%% file: article.tex
\documentclass[extra, fleqn]{gji}
\usepackage[utf8]{inputenc}
\usepackage{graphicx}
\usepackage{amsmath,amssymb}
\usepackage{xcolor}
\usepackage{longtable}
\usepackage{appendix}
\usepackage{times}
\usepackage{hhline}
\usepackage{enumitem}
\usepackage[normalem]{ulem}

\usepackage[]{hyperref} 
\usepackage[squaren,Gray, cdot]{SIunits}
\definecolor{color1}{rgb}{0.22,0.45,0.70}
\hypersetup{%
  pdfauthor={T. Tassin},%
  pdftitle={Geodynamo modelling of double-diffusive convection},%
  pdfcreator={\LaTeX},%
  colorlinks=True,%
  urlcolor=color1,%
  linkcolor=color1,%
  citecolor=color1%
}

\setlist[enumerate, 1]{label=\textit{(\roman*)}}

\newcommand{\lscale}{\mathcal{D}}

\renewcommand{\mitbf}[1]{\hbox{\mathversion{bold}$#1$}}
\newcommand{\grad}{\mitbf{\nabla}}
\newcommand{\laplacien}{\nabla^2}
\newcommand{\dt}{\partial_t}
\newcommand{\dr}{\partial_r}

\newcommand{\cref}{\bar{\xi}}

\newcommand{\vecr}{\mathbf {\hat{r}}}

\newcommand{\vecz}{\mathbf {\hat{z}}}

\newcommand{\vecu}{\mathbf u}

\newcommand{\vecg}{\mathbf g}

\newcommand{\magf}{\mathbf B}

\newcommand{\powertot}{P^\mathrm{tot}}
\newcommand{\powerT}{P_T}
\newcommand{\powerTrel}{P_T^\%}
\newcommand{\powerC}{P_\xi}
\newcommand{\powerDiff}{D_\mathrm{\nu}}
\newcommand{\powerLor}{D_\mathrm{\eta}}

\newcommand{\rat}{Ra_T}
\newcommand{\rac}{Ra_\xi}

\newcommand{\fluxt}{\mathcal{F}^T}
\newcommand{\fluxc}{\mathcal{F}^\xi}
\newcommand{\sourcet}{h^T}
\newcommand{\sourcec}{h^\xi}

\renewcommand{\volume}{V}
\newcommand{\dV}{\mathrm{d}V}
\newcommand{\volumei}{\volume_{i}}
\newcommand{\integralo}{\int_{\volume_o}}

\newcommand{\intoi}{\int_{\volume_o + \volume_i}}

\newcommand{\fdip}{f_\mathrm{dip}}
\newcommand{\rol}{Ro_L}

\newcommand{\ekrel}{\zeta}
\newcommand{\lpic}{\widehat{\ell}}
\renewcommand{\phi}{\varphi}

\begin{document}
\bibliographystyle{gji}
\title[Double-diffusive geodynamo models]{Geomagnetic semblance and 
dipolar-multipolar transition in top-heavy double-diffusive geodynamo models}
\author[T. Tassin, T. Gastine and A. Fournier]
{Théo Tassin$^1$, Thomas Gastine$^1$ and Alexandre Fournier$^1$\\
$^1$Universit\'e de Paris, 
Institut de Physique du Globe de Paris, 
CNRS,
F-75005
Paris, France
}
\maketitle
\input{abstract}


\input{intro}
\input{model}
\input{seuil}
\input{earth_like}
\input{dipole_multipole}
\input{discussion}
\section{Acknowledgments}
\input{acknowledgments}
\section{Data availability}
The data underlying this article will be shared on reasonable request to the
corresponding author.

\appendix
	\onecolumn
	\input{appendix_time}
	\input{table.tex}
	\newpage
	\input{appendix_simu}

\end{document}

%% file: abstract.tex
\begin{abstract} 
Convection in the liquid outer core of the Earth is driven by thermal and chemical perturbations. The main purpose of this study is to examine the impact of double-diffusive convection on  magnetic field generation by means of  three-dimensional global geodynamo models, in the so-called ``top-heavy'' regime of double-diffusive convection, when both thermal and compositional background gradients are destabilizing. Using a linear eigensolver, we begin by confirming that, compared to the standard single-diffusive configuration,  the onset of convection is facilitated by the addition of a second buoyancy source. We next carry out a systematic parameter survey by performing $79$ numerical dynamo simulations. We show that a good agreement between simulated magnetic fields and the geomagnetic field can be attained for any partitioning of the convective input power between its thermal and chemical components. On the contrary, the transition between dipole-dominated and multipolar dynamos is found to strongly depend on the nature of the buoyancy forcing. Classical parameters expected to govern this transition, such as the local Rossby number -a proxy of the ratio of inertial to Coriolis forces- or the degree of equatorial symmetry of the flow, fail to capture the dipole breakdown.  A scale-dependent analysis of the force balance instead reveals that the transition occurs when the ratio of inertial to Lorentz forces at the dominant length scale reaches $0.5$, regardless of the partitioning of the buoyancy power.  The ratio of integrated kinetic to magnetic energy $E_k/E_m$ provides a reasonable proxy of this force ratio.  Given that $E_k/E_m\approx 10^{-4} - 10^{-3}$ in the Earth's core, the geodynamo is expected to operate far from the dipole-multipole transition.  It hence appears that the occurrence of geomagnetic reversals is unlikely related to dramatic and punctual changes of the amplitude of inertial forces in the Earth's core, and that another mechanism must be sought.
\end{abstract}

%% file: intro.tex
\section{Introduction}
The exact composition of Earth's core remains unclear but it is admitted that
it is mainly composed of iron and nickel with a mixture of lighter elements in
liquid state, such as silicon or oxygen \citep[see][for a review]{Hirose2013}.
The ongoing crystallization of the inner core releases light elements and
latent heat at the inner-core  boundary (ICB), while the mantle extracts
thermal energy from the outer core at the core-mantle boundary (CMB). This
combination of processes is responsible for the joint presence of thermal and
chemical inhomogeneities within the outer core.  
 
 Convection with two distinct sources of mass anomaly is termed
 \emph{double-diffusive} convection \citep[e.g.][]{Radko2013}.  A key physical
 parameter of double-diffusive convection is the Lewis number~$Le$, defined as
 the ratio of the thermal diffusivity $\kappa_T$  to the chemical diffusivity
 $\kappa_\xi$.  In the liquid core of terrestrial planets,  $Le$ reaches at
 least $10 - 10^4$ \citep[e.g.][]{Loper1981,Li2000}. 

\begin{figure}
	\centering
	\includegraphics[width=\linewidth]{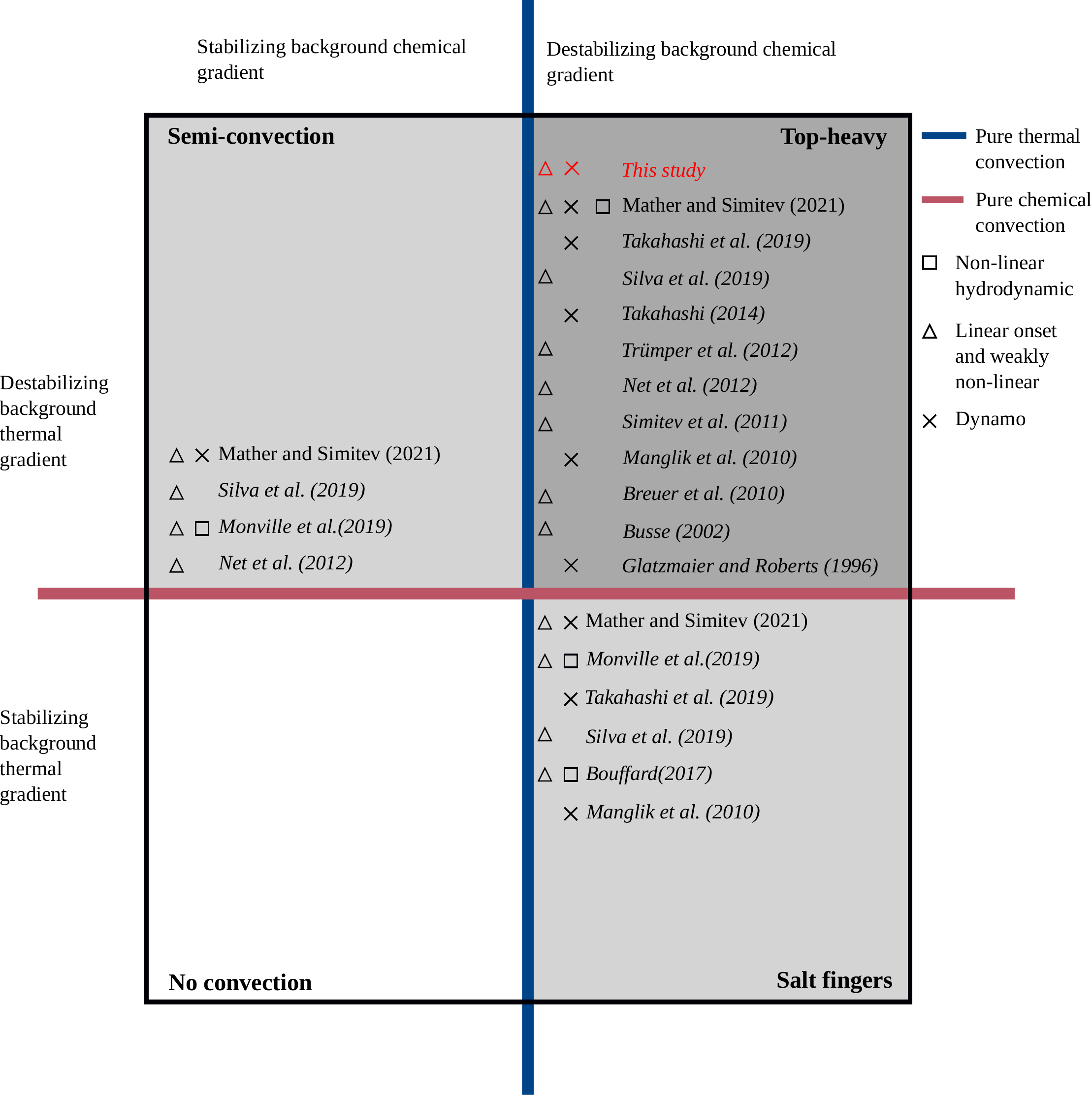}
	\caption{Schematic overview of previous studies on 
		double-diffusive convection and dynamo action
		in terrestrial interiors, inspired by  
		 the regime diagram of \cite{Ruddick1983}.
		 The parameter space has two independent directions, 
		 defined by the prescribed 
		  background temperature and composition
		  gradients, respectively. 
		It is divided into four quadrants, since each 
		 gradient can have a stabilizing or destabilizing
		 effect on fluid motion. 
		 The vertical and horizontal straight lines correspond to purely 
		 thermal and purely chemical convection, respectively 
		 (in which case there exists a unique destabilizing 
		 background profile). 
		The publications that appear in the three quadrants 
		of dynamical interest are the following: \cite{Glatzmaier1996,Busse2002,
		Breuer2010,Manglik2010,Simitev2011,Net2012,
		Trumper2012,Takahashi2014,Bouffard2017,Monville2019,
		Silva2019,Takahashi2019,mather2021}. Crosses denote dynamo studies, triangles
		 linear and weakly non-linear hydrodynamic studies, and 
	 squares non-linear hydrodynamic studies.}
	\label{turner}
\end{figure}

Under double-diffusive conditions, the increase of the background density with
depth does not necessarily imply the stability of the fluid in response to
perturbations.  As shown in~Fig.~\ref{turner}, three configurations can be
considered: 
\begin{enumerate}
	\item the \emph{salt fingering} regime,  when the background 
	  thermal gradient $\grad T_0$ is stabilizing and the mean compositional 
	  gradient $\grad \xi_0$  is destabilizing; 
	\item the \emph{semi-convection} regime, when $\grad T_0$ is destabilizing
	  and $\grad \xi_0$ is stabilizing;
	\item the \emph{top heavy convection} regime (also known as 
	  double-buoyant), when both  $\grad T_0$ and 
	  and $\grad \xi_0$ are destabilizing. 
\end{enumerate}
These three cases correspond to three quadrants in Fig.~\ref{turner}, which is
inspired by \cite{Ruddick1983}.  Note that being located inside one of these
three quadrants does not necessarily guarantee that convection occurs, since,
for all configurations, the onset does not coincide with the origin of the
diagram: a critical contrast in temperature, or composition, or both,  is
required to trigger convective motions.  In addition, this diagram does not
account for the influence of background rotation \citep[see][for the impact 
of rotation on the onset in the salt fingers regime]{Monville2019}, 
or a magnetic field, on the onset . 

 Double-diffusive convection has been extensively studied in physical
 oceanography \citep[e.g.][]{Radko2013}, as heat and salinity provide two
 different sources of buoyancy for seawater, as well as in stellar interiors
 \citep[e.g.][]{Spiegel1972}, where the molecular diffusivity of composition is
 only a tiny fraction of the diffusivity of temperature \citep[e.g.][and
 references therein]{Moll2016}.  To our knowledge, a limited number of studies
 have been devoted to the analysis of double-diffusive convection in the
 context of Earth's core, or more generally in the context of the metallic core
 of terrestrial planets.  They are listed in Fig.~\ref{turner}, with possibly
 several occurrences of a given study when several of the aforementioned
 regimes were considered.  It appears that the top-heavy configuration has been
 investigated the most; this is also the configuration we shall focus on in
 this work. 

Previous studies devoted to the top-heavy regime are listed in the top-right
quadrant of Fig.~\ref{turner}, starting with hydrodynamic, non-magnetic
studies. Considering the $Le\gg1$ limit for a fluid filling a rapidly-rotating
annulus, \cite{Busse2002} theoretically predicted that the addition of chemical
buoyancy facilitated the onset of standard thermal convection, in two ways.
First by lowering the critical value of the Rayleigh number required to trigger
the classical spiralling thermal Rossby waves of size and frequency
proportional to $E^{1/3}$ and $E^{-2/3}$, respectively, where $E=\nu/\Omega
\mathcal{D}^2$ is the non-dimensional Ekman number, $\nu$ being the kinematic
viscosity, $\Omega$ the rotation rate and $\mathcal{D}$ the size of the fluid
domain. Second, and more importantly, by enabling a second class of
instability. The latter is characterized by a much lower onset, independent of
$E$,  a critical length scale of the order of the size of the fluid domain, and
a very small frequency, proportional to $E$. \cite{Busse2002} referred to this
class of instability as ``nearly steady" (i. e. slow) convection,  and he
argued that it could facilitate ``immensely" convection in Earth's core, though
with several caveats \citep[see][for details]{Busse2002}.  \cite{Simitev2011}
pushed the analysis further for various $Le$, and stressed that the critical
onset curve for convective instability in a rotating annulus forms disconnected
regions of instabilities in the parameter space. In particular, the second
family of modes (i.e. the slow modes) are stable whenever the compositional
gradient is destabilizing.  \cite{Trumper2012}, \cite{Net2012} and
\cite{Silva2019} studied the onset of double-diffusive convection in a
spherical shell geometry.  In the top-heavy regime with $3\leq Le \leq 10$,
\cite{Trumper2012} confirmed that the addition of a secondary buoyancy source
facilitates the convective onset. \cite{Net2012} showed that the properties of
the critical onset mode, such as its drift frequency or its azimuthal wave
number, strongly depends on the fractioning between thermal and compositional
buoyancy. Using a linear eigensolver, \cite{Silva2019} carried out a systematic
survey of the onset of convection in spherical shells for the different
double-diffusive regimes. In the top-heavy configuration with $Le=25$, they
showed that the convection onset is characterized by an abrupt change between
the purely thermal and the purely compositional eigenmodes depending on the
relative proportion of the two buoyancy sources. In addition, they demonstrated
that the onset mode features the same asymptotic dependence on the Ekman number
as classical thermal Rossby waves over the entire top-heavy regime (top-right
quadrant of Figure~\ref{turner}), thereby casting some doubt on the likelihood
of the occurrence of slow modes.

\cite{Trumper2012} performed a series of non-linear, moderately supercritical,
rotating convection calculations at constant $Le=10$,  for different
proportions of chemical and thermal driving. To that end, they conducted a
parameter survey, varying the chemical and thermal Rayleigh numbers, to be
defined below, while keeping their sum constant.  This way of sampling the
parameter space, however, does not guarantee that the total buoyancy input
power stays constant. This complicates the interpretation of their results, for
instance regarding the influence of compositional and thermal forcings on the
convective flow properties. 

Let us now turn our attention to the few self-consistent dynamo calculations in
the top-heavy regime published to date (marked with a cross in the top-right
quadrant of Fig.~\ref{turner}). The first integration was reported by
\cite{Glatzmaier1996}. This calculation was an anelastic, double-diffusive
extension of the celebrated Boussinesq simulation of the geodynamo by
\cite{Glatzmaier1995}.  \cite{Glatzmaier1996} assumed  enhanced and equal
values of the diffusivities, i.e. $Le=1$. Consequently, that simulation did not
exhibit stark differences with the purely thermal convective model, except that
the dipole did not reverse over the course of the simulated $40,000$~years. The
second numerical investigation of geodynamo models driven by top-heavy
convection  was conducted by \cite{Takahashi2014}. His models were based on the
Boussinesq approximation with $Le=10$ and relatively large Ekman numbers
($E\geq 2\times 10^{-4}$ ).  In addition, \cite{mather2021} identified a few
top-heavy Boussinesq dynamos with $Le=25$ and $E=10^{-4}$, that used
stress-free mechanical boundary conditions and appeared to be close to onset.
Top-heavy dynamo simulations were also performed by \cite{Manglik2010} and
\cite{Takahashi2019} in the context of modelling Mercury's dynamo.

Double-diffusive models of the geodynamo are the exception rather than the
rule, essentially on the account of Occam's razor. Efforts carried out in the
community since the mid nineteen-nineties have been towards understanding the
most salient properties of the geomagnetic field using a minimum number of
ingredients \citep[e.g.][for a review]{Wicht2019}. To that end, the {\em
codensity} formalism introduced by \cite{Braginsky1995} is particularly
attractive: it assumes  that the molecular values of $\kappa_\xi$ and
$\kappa_T$ can be replaced by a single turbulent transport property.
Consequently, the mass anomaly field can be described by a single scalar,
termed the codensity, that aggregates the two sources of mass anomaly.  This
approach has the benefit of (\textit{i}) removing one degree of freedom and
(\textit{ii}) mitigating the numerical cost by suppressing the scale separation
between chemical and thermal fields when $Le\gg 1$.  With regard to
Fig.~\ref{turner}, this amounts to restricting the diagram to either the
vertical or the horizontal straight line. 

The codensity formalism was quite successful in reproducing some of the best
constrained features of the geomagnetic field and its secular variation
\cite[e.g.][]{Christensen2010,Aubert2013,Schaeffer2017,Wicht2019}.  Most
geodynamo models actually assume that the diffusivity of the codensity field
equals the kinematic viscosity, yielding a Prandtl number of unity.  A
remarkable property of the geodynamo that remains to be explained
satisfactorily from the numerical modelling standpoint, is its ability to
reverse its polarity every once in a while, that is to go from a
dipole-dominated state to another dipole-dominated state through a transient
multipolar state \citep[see e.g.][for a recent review of the relevant
paleomagnetic data]{Valet2016}. A possibility is that the geodynamo has been,
at least punctually in its history, in a dynamical state that can enable the
switch between dipole-dominated and multipolar states to occur.  A key question
that follows is therefore: what are the physical processes that control the
transition between dipole-dominated dynamos and multipolar dynamos?  This has
been analyzed intensively numerically, starting with the systematic approach of
\cite{Kutzner2002}, who demonstrated that a stronger convective driving led to
a dipole breakdown, and that for intermediate values of the forcing, the
simulated field could oscillate between dipolar and multipolar states. 

\cite{Sreenivasan2006} showed that an increasing role of inertia (through
stronger driving) perturbed the dominant Magneto-Archimedean-Coriolis force
balance to the point that it led to a less structured and less dipole-dominated
magnetic field.  In the same vein, \cite{Christensen2006} assumed that the
transition is due to a competition between inertia and Coriolis force. They
introduced a diagnostic quantity termed the \emph{local Rossby number}
$Ro_L=u_\mathrm{rms}/\Omega L$, as a proxy of this force ratio. Here
$u_\mathrm{rms}$ denotes the average flow speed and $L$ is an integral measure
of the convective flow scale. Based on their ensemble of simulations,
\cite{Christensen2006} concluded that the breakdown of the dipole occurred
above a critical value of about $Ro_L\simeq 0.12$, in what appeared a
relatively sharp transition \citep[see also][]{Christensen2010a}. If this
reasoning gives a satisfactory account of the numerical dataset, its
extrapolation to Earth's core regime raises questions
\citep[e.g.][]{Oruba2014}.  Since the geomagnetic dipole reversed in the past,
the numerical evidence collected so far \citep[see][for a review]{Wicht2010}
suggests that the geodynamo could lie close to the transition between dipolar
and multipolar states.  This implies that $Ro_L$ could be of the order of $0.1$
for the Earth's core.  Geomagnetic reversals should then reflect the action of
a convective feature of scale $L$ of about $\unit{50}{\meter}$
\citep[see][]{Davidson2013,Aubert2017}.  It is very unlikely that such a
small-scale flow could significantly alter a dipole-dominated magnetic field. 

According to \cite{Soderlund2012}, the breakdown of the dipole is rather due to
a decrease of the relative helicity of the flow.  In numerical dynamo
simulations, coherent helicity favors large-scale poloidal magnetic field
through the $\alpha$-effect \citep[see][]{Parker1955} at work in convection
columns and it therefore contributes actively to the production and the
maintenance of dipolar field \citep[e.g.][]{Olson1999}.  Conversely, the
dipolar field can promote a more helical flow, with the Lorentz force enhancing
the flow along the axis of convection columns, as shown by
\cite{Sreenivasan2011}.   By measuring the integral force balance for dipolar
and multipolar numerical dynamos, \cite{Soderlund2012} noticed that the
Coriolis force remained dominant, even in multipolar models.  Accordingly, they
suggested that, in their models,  the modification of the flow structure was
rather controlled by a competition of second-order forces, with the ratio of
inertia to viscous forces as the parameter controlling the transition.  

The role played by viscous effects was further stressed by \cite{Oruba2014},
who proposed that the transition from dipolar to multipolar dynamos in the
numerical dataset was controlled by a triple force balance between Coriolis
force, viscosity, and inertia.  A local Rossby number constructed using the
viscous lengthscale, $E^{1/3}\,\mathcal{D}$, as opposed to the integral scale
$L$ discussed above, carries the same predictive power in separating dipolar
from multipolar dynamo models (their Fig.~4).  This argument was subsequently
refined by \cite{Garcia2017}, who included the Prandtl number dependence of the
critical convective length scale when defining the local Rossby number. From a
mechanistic point of view, \cite{Garcia2017} showed that the dipole breakdown
was not necessarily correlated to a decrease of the relative helicity, but
rather to a weakening of the equatorial symmetry of the flow.  Introducing the
proportion of kinetic energy contained in this equatorially symmetric
component, they demonstrated that the transition could  be satisfactorily
explained by a sharp decrease of that quantity when the refined local Rossby
number exceeded a value of $0.2$ (their Fig.~3d). This would imply that the
transition from a dipolar state to a multipolar state would essentially be a
hydrodynamic transition. Discussing the implication of their results for the
geodynamo, they clearly stated that the role of inertia was presumably
overestimated in the numerical dataset that had been investigated so far,
stressing the need for stronger field dynamos, where the magnetic field could
possibly have an active role. 

The early numerical dataset admittedly contained a majority of dynamos
operating at large Ekman numbers, and relatively low magnetic Prandtl number
$Pm$. In the dynamos studied by \cite{Soderlund2012}, the convective flow was
not dramatically altered by the presence of a self-sustained magnetic field.
Since then, a large number of simulations have been published with lower values
of $E$ and comparatively larger values of $Pm$
\citep{Yadav2016,Schaeffer2017,Schwaiger2019,Menu2020}.  For those strong-field
dynamos, in the sense of a ratio of bulk magnetic energy to bulk kinetic energy
larger than one, the magnetic field has a significant impact on the flow, and
on the dipolar-multipolar transition. \cite{Menu2020} reported simulations with
a prevailing Lorentz force, that remain dipolar way beyond the supposedly
critical $Ro_L\simeq 0.12$ value.  Given that Earth hosts a strong-field
dynamo, it is then worth investigating whether the competition between Lorentz
force and inertia may actually lead to the transition. 

In order to shed light on this particular issue, and to further strengthen our
understanding of double-diffusive dynamos, we have performed a suite of $79$
novel dynamo models, comprising $44$, $20$, $15$ simulations with top-heavy,
purely thermal, purely chemical driving, respectively.  The goal of this work
is twofold: on the one hand, we will follow an approach similar to that of
\cite{Takahashi2014} and study to which extent the relative proportion of
chemical to thermal driving impacts the earth-likeness (in a morphological
sense) of the simulated magnetic fields. On the other hand, we will examine
whether this proportion has an impact on the dipolar to multipolar transition.
\cite{Takahashi2014} reported a drop of the dipolarity when the relative
contribution of thermal convection to the total input power exceeds $65\,\%$.
We will analyze whether this was a fortuitous consequence of the cases he
considered, and whether we can find a more general rationale to explain the
transition, by careful inspection of the force balance at work. 

This paper is organized as follows: the derivation of the governing equations
and their numerical approximations are presented in Section~\ref{model}. The
results are described in Section~\ref{results}.  We proceed by firstly
investigating the onset of convection for top-heavy convection.  The impact of
the input power distribution on the Earth-likeness is then explored. Finally we
examine the transition between dipolar and multipolar dynamos.
Section~\ref{discussion} discusses the results and their geophysical
implications.

%% file: model.tex
\section{Model and methods}
\label{model}
\subsection{Hypotheses}
\label{hp}
We operate in spherical coordinates $(r,\theta,\varphi)$ and consider a
spherical shell of volume $V_o$ filled with a fluid delimited by the inner core
boundary (ICB), located at the radius $r_i$, on one side and by the core mantle
boundary (CMB), located at the radius $r_o$, on the other side with $r_i/r_o =
0.35$.  The shell rotates about the $\vecz$-axis with a constant rotation rate
$\Omega$, where $\vecz$ is the unit vector in the direction of rotation.  The
equation of state 
\begin{equation}
	\rho = \rho_0 [1-\alpha_T(T - T_0) -\alpha_\xi(\xi - \xi_0)],
\end{equation}
describes how the density of the fluid $\rho$ varies with temperature $T$ and
composition $\xi$.  In this equation, $\alpha_T$ and $\alpha_\xi$ are the
coefficients of thermal and chemical expansion, $T_0$, $\rho_0$ and $\xi_0$ the
average temperature, density and composition of lighter elements in the outer
core.

The properties of the fluid, including its kinematic viscosity $\nu$, its
magnetic diffusivity $\eta$, its specific heat $C_p$, its chemical and thermal
diffusivity ($\kappa_\xi, \kappa_T$), its coefficients of thermal and chemical
expansion ($\alpha_T$, $\alpha_\xi$) are assumed to be spatially uniform and
constant in time.  Due to the almost uniform density in the outer core, we also
assume a linear variation of the acceleration of gravity $\vecg$ with radius.
The physical and thermodynamical properties of the Earth's core relevant for
this study are given in Table~\ref{phy_param}.

\begin{table*}
	\caption{Physical and thermodynamical parameters of Earth's  
		 outer core relevant for this study. The corresponding references
		 are listed in the rightmost column, which may comprise several 
		 entries if a bracket of values are provided. 
		 }
	\label{phy_param}
	\begin{tabular}{l l l l}\hline
		\textbf{Definition} &  \textbf{Symbol} & \textbf{Value} 
		& \textbf{Reference} (Lower bound - Upper bound) \\\hhline{====}
		Inner radius & $r_i$ & $\unit{1221.5}{\kilo\meter}$  
		& \cite{Dziewonski1981}\\
		Outer radius & $r_o$ & \unit{3480}{\kilo\meter} & \cite{Dziewonski1981}\\
		Earth angular velocity & $\Omega$ & \unit{7.29 \times 10^{-5}}
		{\rad \usk \reciprocal \second}	&\\
		Gravitational acceleration at CMB & $g_o$ & \unit{10.68}
		{\meter \usk \rpsquare \second}	& \cite{Dziewonski1981}\\
		Core density at CMB & $\rho_o$ & \unit{9 \times 10^3}
		{\kilogram \usk \rpcubic \meter} & \cite{Dziewonski1981}\\	
		Specific heat &	$C_p$ &	\unit{(850 \pm 80)}{\joule \usk 
		\reciprocal \kilogram \usk \reciprocal \kelvin}	&
		\cite{Labrosse2003}\\
		Heating power from core & $Q$ & \unit{6 -16}{\tera \watt} 
		& \cite{Buffett2015}\\
		Thermal conductivity at CMB & $k_T$ & $25 - 100$ 
		$\watt \usk \reciprocal \meter \usk \reciprocal \kelvin$ & 
		\cite{Konopkova2016} - \cite{Pozzo2013,Zhang2020}\\
		Thermal diffusivity & $\kappa_T$ & ($0.3 - 1.4$) 
		$\times 10^{-5}$ $\squaren \meter \usk \reciprocal \second$ & 
		Estimated using values of $k_T$, $\rho_o$ and $C_p$.\\
		Coefficient of thermal expansion & $\alpha_T$ &	
		\unit{(1.3 \pm 0.1) \times 10^{-5}}{\reciprocal \kelvin} 
		& \cite{Labrosse2003}\\
		Kinematic viscosity & $\nu$ & $\unit{10^{-6}}{\squaren \meter 
		\usk \reciprocal \second} $ & \cite{Roberts2013}\\
		Magnetic diffusivity & $\eta$ &	\unit{0.5 - 2.9}
		{\squaren \meter \usk \reciprocal \second} & \cite{Pozzo2013} 
		- \cite{Konopkova2016}\\
		Estimated magnetic field strength & $B_\mathrm{rms}$ &
		$\unit{4 \times 10^{-3}}{\tesla}$ & \cite{Gillet2010} \\
		Superadiabatic composition contrast & $\Delta \xi$ 
		& 0.02 - 0.053 & \cite{Badro2007} - \cite{Anufriev2005} \\
		Chemical diffusivity & $\kappa_\xi$ & ($3 \times 10^{-9}$ -- 
		$4.2 \times 10^{-7}$) $\squaren \meter \usk \reciprocal \second$  
		& \cite{Loper1981} - \cite{Li2000} \\
		Coefficient of chemical expansion & $\alpha_\xi$ &
		$0.6 - 0.83$ & \cite{Braginsky1995} - \cite{Labrosse2015}\\
		Estimated flow velocity & $u_\mathrm{rms}$  & $\unit{(0.3 - 2.0) 
		\times 10^{-3}}{\meter \usk \reciprocal \second}$ & \cite{Finlay2011}\\
	\end{tabular}
	
        \caption{Dimensionless control parameters. The two rightmost columns 
provide estimates of these parameters for Earth's core and the values 
spanned by the simulations computed in this study. Earth's core values 
were estimated thanks to Tab.~\ref{phy_param}.}
        \label{dimensionless}
        \begin{tabular}{l l l  l l}\hline
        \textbf{Name} & \textbf{Symbol} & \textbf{Definition} & \textbf{Core} 
	& \textbf{This study}\\ \hhline{=====}
	Ekman & $E$ & $\nu/\Omega \lscale^2$ & $10^{-15}$ & $10^{-4}$ -- $10^{-5}$\\
	Thermal Rayleigh & $\rat$  & $\alpha^T g_0 \lscale^4 \fluxt_i/\nu 
	\kappa_T$ &  $10^{26} - 10^{28}$ & $10^6 - 10^{10}$\\
        Chemical Rayleigh & $\rac$ & $\alpha^\xi g_0 \lscale^4 \fluxc_i/\nu 
	\kappa_\xi$ & $10^{30} - 10^{33}$ & $10^7 - 10^{12}$\\
        Magnetic Prandtl & $Pm$ & $\nu/\eta$ & $(3.4-20) \cdot 10^{-7}$ & 0.5 -- 5\\
        Thermal Prandtl & $Pr$ & $\nu/\kappa_T$ & $0.08$ -- $0.25$ & 0.3\\
        Schmidt & $Sc$ &$\nu/\kappa_\xi$ & $2$ -- $300$ & 3\\
	Lewis & $Le$ &$\kappa_T/\kappa_\xi$ & $9$ -- $4000$ & 10\\
        \end{tabular}
	\caption{Characteristic time scales for the Earth's core. The values were 
		 estimated thanks to Tab.~\ref{phy_param}.}
        \label{time}
        \begin{tabular}{l l l l }\hline
		\textbf{Name} & \textbf{Symbol} & \textbf{Definition} &   
		\textbf{Core}\\ \hhline{====}
	Typical rotation time 
	& $\tau_\Omega$ & $1/\Omega$ & \unit{4}{\hour} \\
	Turnover time & $\tau_\mathrm{adv}$ & $\lscale/u_\mathrm{rms}$ & 
	\unit{30 - 250}{yr} \\
	Magnetic diffusion time & $\tau_\eta$ & $\lscale^2/\eta$ & 
	\unit{10^5 - 10^6}{yr} \\
	Thermal diffusion time & $\tau_T$ & $\lscale^2/\kappa_T$ & 
	\unit{10^9 - 10^{10}}{yr} \\
	Viscous diffusion time & $\tau_\nu$ & $\lscale^2/\nu$ & 
	\unit{10^{11}}{yr} \\
	Chemical diffusion time & $\tau_\xi$ & $\lscale^2/\kappa_\xi$ 
	& $10^{11}$ -- \unit{10^{13}}{yr}\\
        \end{tabular}

	\caption{Output parameters of the numerical simulations and their 
	         estimates for Earth's core}	
	\label{output}
	\begin{tabular}{l l l l l p{3cm}l}\hline
		\textbf{Name} & \textbf{Symbol} & \textbf{Definition} &	
		\textbf{Earth's core} & \textbf{This study} & 
		\textbf{Reference}\\ \hhline{======}
		Relative thermal convective power & $\powerT^\%$ & 
		see Eq.~\ref{powert} & $20 - 70\,\%$ & $0 - 100$ $\%$ & 
		\cite{Lister1995} - \cite{Takahashi2014}\\
		Rossby & $Ro$ & $u_\mathrm{rms} /\Omega \lscale$ & $(1.7 - 12) 
		\times 10^{-6}$ & $0.002 - 0.1$ & Table~\ref{phy_param}\\
		Local Rossby & $Ro_L$ &	$u_\mathrm{rms}/ \Omega L$ & 
		$4.7 \times 10^{-5} - 0.09$ & $0.009 - 0.45$ & 
		\cite{Davidson2013} - \cite{Olson2006}\\				
		Relative equat. symmetric kinetic energy & $\ekrel$ & &	
		$0.78 - 0.9$ & $0.65 - 0.96$  & \cite{Aubert2017}\\
		Magnetic Reynolds & $Rm$ & $u_\mathrm{rms} \lscale/\eta$ & 
		$(0.2 - 9) \times 10^3$ & $10^2 - 6 \times 10^3$& Table~\ref{phy_param}\\
		Elsasser & $\Lambda$ & $B_\mathrm{rms}^2/ \mu_0 \eta \rho_o 
		\Omega$ & $6.7 - 39$	& $0.3 - 3 \times 10^2$	 & 
		Table~\ref{phy_param}\\
		Dipolarity parameter &$f_\mathrm{dip}$ & & $0.6 - 0.7$ & 
		$0.1 - 1$ & \cite{Gillet2015}
	\end{tabular}
\end{table*}
\subsection{Governing equations}
\label{gov_eq}
Convection of an electrically-conducting fluid gives rise to a magnetic field
$\magf$.  The state of the fluid is then described by the velocity field
$\vecu$, the magnetic field $\magf$, the pressure $p$, the temperature $T$ and
the composition $\xi$.  The equations governing the dynamics of the flow under
the Boussinesq approximation are cast in a non-dimensional form.  We adopt the
thickness of the shell $\lscale = r_o - r_i$ as reference length scale and the
viscous diffusion time $\lscale^2/\nu$ as time scale.  A velocity scale is then
given by $\nu/\lscale$.  Composition is scaled by $\partial \xi = |\dr \xi
(r_i)|\lscale$, pressure by $\rho_0 \left(\nu/\lscale \right)^2$, power by
$\nu^3 \rho_0 /\lscale$ and magnetic induction by $\sqrt{\rho_0 \mu_0 \eta
\Omega}$, where $\mu_0$ is the magnetic permeability of vacuum. Temperature
unit is based on the temperature gradient at $r_i$, $\partial T = |\dr T
(r_i)|\lscale$, or at $r_o$, $\partial T = |\dr T (r_o)|\lscale$, depending on
the thermal boundary conditions. Under the Boussinesq approximation, the
equation for the conservation of mass is 
\begin{equation}
   \grad \cdot \vecu = 0.
   \label{mass_conservation}
\end{equation}
The dynamics of the flow is described by the Navier-Stokes equation, 
expressed in the frame rotating with the mantle 
\begin{equation}
	\begin{array}{c@{-}l}
		\dt \vecu + \vecu \cdot\grad\vecu  =
			&	\dfrac{2}{E} 
			 	\vecz \times \vecu + 
				\left(\dfrac{\rat}{Pr}T + 
			 	\dfrac{\rac}{Sc} \xi \right) 
			 	\dfrac{r}{r_o}\vecr \\[0.5cm]
			&	\grad p + \dfrac{1}{EPm} 
				[(\grad \times \magf) \times \magf]   + \grad^2 \vecu ,
	\end{array}
  \label{NS_dim}
\end{equation}
where $\vecr$ is the unit vector in the radial direction. The time evolution
of the magnetic field under the magnetohydrodynamics approximation is given by
the induction equation

\begin{equation}
	\dt \magf = \grad \times (\vecu \times \magf) + 
\dfrac{1}{Pm} \grad^2 \magf \quad \mathrm{with} \quad \grad \cdot \magf = 0. 
  \label{induction}
\end{equation}
Finally, the evolution of entropy  and composition are governed by the similar
transport equations
\begin{equation}
	\dt T + \vecu \cdot \grad T = \dfrac{1}{Pr} \laplacien T + \sourcet,
	\label{thermal_transport}
\end{equation}
and
\begin{equation}
    \dt \xi  + \vecu \cdot \grad \xi = \dfrac{1}{Sc} \laplacien \xi + \sourcec,
    \label{chemical_transport}
\end{equation}
where $\sourcet$ is a volumetric internal heating  and $\sourcec$ a chemical
volumetric source.

The set of equations~(\ref{mass_conservation} -- \ref{chemical_transport}) is
governed by 6 dimensionless numbers: The Ekman number $E$ expresses the ratio
between viscous and Coriolis forces 
\begin{equation*}
	E = \dfrac{\nu}{\Omega \lscale^2},
\end{equation*}
the Prandtl (Schmidt) number between viscous and thermal (chemical) 
diffusivities 
\begin{equation*}
	Pr = \dfrac{\nu}{\kappa_T} \quad \mathrm{and} \quad Sc = \dfrac{\nu}{\kappa_\xi}
\end{equation*}
and the magnetic Prandtl number between viscous and magnetic diffusivities 
\begin{equation*}
	Pm = \dfrac{\nu}{\eta}.
\end{equation*}
The thermal and chemical Rayleigh numbers
\begin{equation*}
	\rat = \dfrac{\alpha_T g_0 \lscale^3 \partial T}{\nu \kappa_T} \quad 
	\mathrm{and} \quad 
	\rac = \dfrac{\alpha_\xi g_0 \lscale^3 \partial \xi}{\nu \kappa_\xi},
\end{equation*}
where $g_o$ is the gravitational acceleration at the CMB, measure the vigour of
thermal and chemical convection.  Note that the Lewis number $Le$ discussed in
the Introduction is the ratio of the Schmidt number to the Prandtl number, 
\begin{equation*}
	Le = \dfrac{\kappa_T}{\kappa_\xi} = \frac{Sc}{Pr}. 
\end{equation*}

Table~\ref{dimensionless} provides estimates of these control parameters for
the Earth's core.  These nondimensional number express the ratio of
characteristic physical time scales
\begin{equation*}
	E = \dfrac{\tau_\Omega}{\tau_\nu},\ Pm = \dfrac{\tau_\eta}{\tau_\nu}\quad \mathrm{and}\quad Le = \dfrac{\tau_\xi}{\tau_T},
\end{equation*}
where $\tau_\Omega = 1/\Omega$ is the typical rotation time, $\tau_\nu =
\lscale^2/\nu$ the viscous diffusion time, $\tau_\eta = \lscale^2/\eta$ the
magnetic diffusion time, $\tau_T = \lscale^2/\kappa_T$ the thermal diffusion
time and $\tau_\xi = \lscale^2/\kappa_\xi$ the chemical diffusion time.

Earth's outer core evolves on a broad range of time scales, as one can glean
from the inspection of Table~\ref{time}.  In particular,  even if the evolution
of temperature and composition are governed by similar transport equations (see
Eq.~\ref{thermal_transport} and \ref{chemical_transport}), thermal diffusion is
much more efficient than chemical diffusion,  which causes the Lewis number to
greatly exceeds unity (see Tab.~\ref{dimensionless}).  This implies that the
typical lengthscale of chemical heterogeneities is possibly several orders of
magnitude smaller than the length scale of thermal heterogeneities.  The values
of the control parameters spanned by the simulations presented in this work are
discussed in section~\ref{sec:explo}. 

\subsection{Boundary conditions}
\label{boundary}
\label{sec:bc}
The inner core is growing and is ejecting light elements because of its
crystallization, which in principle yields coupled boundary conditions for
temperature and composition \citep[e.g.][]{Glatzmaier1996, Anufriev2005}.  For
the sake of simplicity, thermal and chemical boundary conditions are considered
as decoupled here, as in e.g. \cite{Takahashi2014}.  We consider two different
setups to explore the impact of varying the thermal boundary conditions:
\begin{enumerate}
	\item Fixed fluxes: thermal and composition fluxes are imposed at both 
boundaries with
\begin{equation}
\left\{
\begin{array}{rll}
\dr T(r_i) = -1,\\
\dr \xi(r_i) = -1,
\end{array}
~
\begin{array}{rll}
\dr T(r_o) = 0,\\
\dr \xi(r_o) = 0.
\end{array}
\right.
\label{fixed_fluxes}
\end{equation}
	\item Hybrid: temperature is fixed at the ICB while the temperature flux is 
imposed at the CMB, the boundary conditions on chemical composition are 
the same as in the previous setup
\begin{equation}
\left\{
\begin{array}{rll}
T(r_i) = 0,\\
\dr \xi(r_i) = -1,
\end{array}
~
\begin{array}{rll}
\dr T(r_o) = -1,\\
\dr \xi(r_o) = 0.
\end{array}
\right.
\label{hybrid}
\end{equation}
\end{enumerate}
No-slip mechanical boundary conditions are used at both boundaries.  The mantle
is assumed to be insulating, such that the magnetic field at the CMB has to
match a source-free potential field.  The inner core is treated as a rigid
electrically-conducting sphere which can freely rotate around the $\vecz$-axis
\citep[e.g.][]{Hollerbach2000,Wicht2002}.  Its rotation is a response to the
viscous and magnetic torques exerted by the outer core on the inner core. The
conductivity of the inner core is assumed to be equal to that of  the outer
core, and its moment of inertia is calculated by using the same density as the
liquid outer core.

\subsection{Numerical approach}
\label{numerical}
We solve the system of equations~(\ref{mass_conservation}--
\ref{chemical_transport}) using the open-source geodynamo code
MagIC\footnote{\url{https://magic-sph.github.io/}} \citep{Wicht2002,
Gastine2016}.  This code has been validated against a benchmark for
double-diffusive convection initiated by \cite{Breuer2010}. 

The solenoidal vectors $\vecu$ and $\magf$ are decomposed in poloidal and
toroidal potentials 
\begin{equation*}
        \left\{
		\begin{array}{l p{0.1cm} l}
			\vecu(\mathbf{r}, t) &=& \grad \times \grad \times \left[ W(\mathbf{r}, 
t)\vecr\right] + \grad \times \left[Z(\mathbf{r}, t)  \vecr\right], \\
			\magf(\mathbf{r}, t)&=& \grad \times \grad \times \left[ G(\mathbf{r}, t)\vecr \right] + \grad \times \left[H(\mathbf{r}, t)\vecr\right], \\
        \end{array}
        \right.
        \label{decomposition}
\end{equation*}
where $\mitbf{r}$ is the radius vector.  The new unknowns are then $W$, $Z$,
$G$, $H$, $T$, $\xi$ and $p$. 

\par Each of these scalar fields is expanded in spherical harmonics  to maximum
degree and order $\ell_\mathrm{max}$ in the horizontal direction.  The
spherical harmonic representation of the magnetic poloidal potential $G$ reads
\begin{equation*}
	G(r, \theta, \phi, t) \simeq \displaystyle \sum_{\ell = 0}^{\ell_\mathrm{max}} \sum_{m = -\ell}^{\ell} G_{\ell m}(r, t) Y^m_\ell(\theta, \phi)
\end{equation*}
where $G_{\ell m}(r, t)$ is the coefficient associated to $Y^m_\ell$, the
spherical harmonic of degree $\ell$ and order $m$.  The non-linear terms are
calculated in physical space.  The open-source SHTns
library\footnote{\url{https://bitbucket.org/nschaeff/shtns}}
\citep[see][]{Schaeffer2013} is used to compute the forward and inverse
spectral transforms on the unit-sphere. 

In the radial direction, MagIC uses either a finite difference scheme, or a
Chebyshev collocation method \citep[see][]{Boyd2001}.  The finite difference
grid, whose number of points is denoted by $N_r$, is regularly spaced in the
bulk of the domain, and follows a geometric progression near the boundaries
\citep[see][]{Dormy2004}.  When the collocation approach is selected, Chebyshev
polynomials are truncated at degree $N$ and the $N_r$ collocation points $r_k$
are defined by
\begin{equation}
	\forall k \in [\![1, N_r]\!], \
	\left\{
	\begin{array}{rrl}
		r_k &=& \dfrac{1}{2}\left[ (r_o-r_i)x_k + r_o + r_i \right]\\[0.3cm]
		x_k &=& \cos \left[\dfrac{(k - 1)\pi}{N_r - 1} \right]. 
	\end{array}
	\right.
\label{Lobatto}
\end{equation}
Due to the particular choice of spatial grid given by the equation above, the
transforms between physical grid and Chebyshev representation are carried out
by fast discrete cosine transform \citep[see][chapter~12]{Press1992}.  This
discretisation yields a point densification close to the boundaries which could
impose severe time step restrictions when the magnetic field is strong
\citep[see][]{Christensen1999}.  To mitigate this effect, we adopt the mapping
proposed by \cite{Kosloff1993} and replace $x_k$ in Eq.~(\ref{Lobatto}) by
\begin{equation*}
	\forall k \in [\![1, N_r] \!],\ X_k = \dfrac{\arcsin(\alpha x_k)}{\arcsin(\alpha)} \quad \mathrm{with} \quad \alpha \in ]0, 1],
\end{equation*}
where $\alpha$ is the mapping coefficient.  To maintain the spectral
convergence of the simulation $\alpha$ has to verify 
\begin{equation}
	\alpha \leq \alpha_\mathrm{max} = \left\{ \cosh 
\left[\dfrac{|\ln(\epsilon)|}{N_r - 1}\right]\right\}^{-1},
\label{eq:alphamax}
\end{equation}
where $\epsilon$ is the machine precision.
\begin{figure}
	 \centering
	 \includegraphics[width=\linewidth]{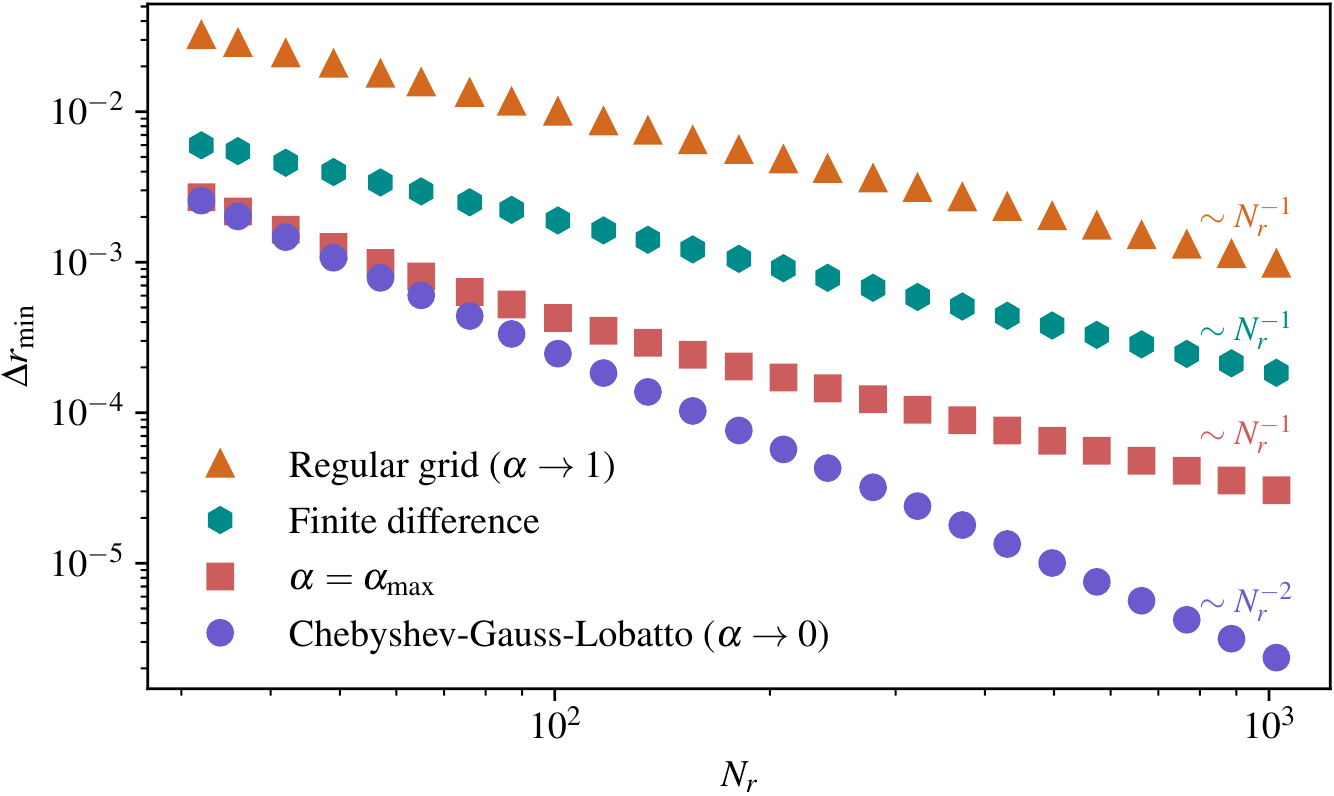}
	 \caption{Minimal grid spacing $\Delta r_\mathrm{min}$ between two radial 
points as a function of the number of collocation points 
$N_r$ for a regularly-spaced grid, the grid used when 
finite differences are employed, the collocation grid with the mapping by 
\cite{Kosloff1993} with a mapping coefficient $\alpha_\mathrm{max}$ 
(Eq.~\ref{eq:alphamax}) and the 
standard Gauss-Lobatto grid.}
	 \label{alpha}
\end{figure}

Figure~\ref{alpha} shows the minimum grid spacing $\Delta r_\mathrm{min}$ as a
function of $N_r$ for a regular grid, the finite difference grid with
geometrical clustering near the boundaries, and two collocation grids, with
$\alpha\rightarrow 0$ (Gauss-Lobatto grid) and $\alpha=\alpha_\mathrm{max}$.
Because $\Delta r_\mathrm{min}\sim N_r^{-2}$ when using the classical
Gauss-Lobatto grid, adopting the mapping by \cite{Kosloff1993} yields a
possible increase of  $\Delta r_\mathrm{min}$ by a factor $2-3$ when $N_r
\gtrsim 100$.  The time step size could in principle rise by a comparable
amount, should it be controlled by the propagation of Alfv\'en waves in the
vicinity of the boundaries \citep{Christensen1999}. Using the finite difference
method enables even larger grid spacing and hence possible additional gain in
the time step size.  This speed-up shall however be mitigated by the fact that
$N_r$ has to be increased by a factor $1.5$ to $3$ when using finite
differences to achieve an accuracy comparable to that of the Chebyshev
collocation method \citep[see][]{Christensen2001, Matsui2016}. 

$G_{\ell m}$ is then expanded in truncated Chebyshev series
\begin{equation}
G_{\ell m}(r_k, t) \simeq \sqrt{\frac{2}{N_r - 1}} 
\displaystyle\sideset{}{''}\sum^{N - 1}_{n = 0} G_{\ell m n}(t) T_n(r_k),
	\label{lmn}
\end{equation}
with 

\begin{equation*}
	G_{\ell m n}(t) \simeq \sqrt{\frac{2}{N_r - 1}} \sideset{}{''}\sum^{N_r}_{n 
= 1} G_{\ell m}(x_k, t) T_n(x_k),
\end{equation*}
where the double primes on the summations indicate that the first and the last 
indices are multiplied by one half \citep[see][]{Glatzmaier1984}. In the above 
equations, $T_n(x_k)$ is the $n$-th order Chebyshev polynomial at the 
collocation point $x_k$ defined by

\begin{equation*}
	T_n(x_k) = \cos[n \arccos(x_k)] = \cos\left[ \frac{\pi n (k - 1)}{N_r -1} \right].
\end{equation*}
Further details on spherical harmonics and Chebyshev polynomials expansion can
be found in \cite{Tilgner1999} and \cite{Christensen2015}.

Once the spatial discretisation has been specified, the set of
equations~(\ref{mass_conservation}--\ref{chemical_transport}) complemented by
the boundary conditions (see Eq.~\ref{fixed_fluxes} and \ref{hybrid}) forms a
semi-discrete system where only the time discretisation remains to be
expressed. As an example, the time evolution of the poloidal potential for the
magnetic field $G_{\ell m}(r_k)$ (see Eq.~\ref{lmn}) can be written as an
ordinary differential equation

\begin{equation}
\left\{
\begin{array}{lcl}
	\dfrac{\mathrm{d}G_{\ell m}}{\mathrm{d} t}(r_k, t) &=& \mathcal{E}[\mathbf{u}, \mathbf{B}] + \mathcal{I}[G_{\ell m}], \\[0.3cm]
	G_{\ell m}(x_k, t_0) &=& G_{\ell m}^0(x_k),
\end{array}
\right.
\label{ordinary}
\end{equation}
where $G_{\ell m}^0(r_k)$ is the initial condition, $\mathcal{E}$ a non
linear-function of $\mathbf{u}$ and $\mathbf{B}$ and $\mathcal{I}$ a linear
function of $G_{\ell m}$.  The above equation serves as a canonical example of
the treatment of the different contributions: the non-linear terms are treated
explicitly (function~$\mathcal{E}$) while the remaining linear terms are
handled implicitly (function~$\mathcal{I}$).  In MagIC, several
implicit/explicit (IMEX) time schemes are employed to time advance the set of
equations (\ref{mass_conservation}--\ref{chemical_transport}) from $t$ to $t +
\delta t$:
\begin{enumerate}
	\item a combination of a Crank-Nicolson for the implicit terms and a 
	      second-order Adams-Bashforth for the explicit terms called CNAB2 
	      \citep[see][]{Glatzmaier1984}, 
	\item two IMEX Runge-Kutta : PC2 \citep[see][]{Jameson1981} and BPR353 
	      \citep[see][]{Boscarino2013}. 
\end{enumerate}		
CNAB2 has been commonly used in geodynamo models since the pioneering work of
\cite{Glatzmaier1984}. IMEX Runge-Kutta schemes have been rarely employed in
the context of geodynamo models \citep{Glatzmaier1996}, rapidly-rotating
convection in spherical shells \citep{Marti2016} or quasi-geostrophic models of
2-D convection \citep{Gastine2019}.  For IMEX Runge-Kutta schemes, $s$
substages are solved to time-advance Eq.~(\ref{ordinary}) from $t$ to $t +
\delta t$ 
\begin{equation}
G_{\ell m}^i(r_k) = G_{\ell m}(r_k, t) 
				  + \delta t \sum_{j=0}^{i} 
				  \left( a_{i,j}^\mathcal{E} \mathcal{E}^j + a_{i,j}^\mathcal{I} \mathcal{I}^j \right),
\end{equation}
where $i \in [\![0,s]\!]$, $G_{\ell m}^i$ is the intermediate solution at substage $i$, $\mathcal{E}^j = \mathcal{E}[\mathbf{B}, \mathbf{u}](t + c_j^\mathcal{E} \delta t)$ and $\mathcal{I}^j = \mathcal{I}[G_{\ell m}](t + c_j^\mathcal{I} \delta t)$. 
$G_{\ell m}(r_k, t + \delta t)$ is then given by
\begin{equation}
	G_{\ell m}(r_k, t + \delta t) = G_{\ell m}(r_k, t) + \delta t \sum_{j=0}^s 
\left(b_j^\mathcal{E} \mathcal{E}^j + b_j^\mathcal{I} \mathcal{I}^j \right)
\label{eq:assembly_imex}
\end{equation} 
In the above equations, the matrices $\mathbf{a}^\mathcal{I}$ and
$\mathbf{a}^\mathcal{E}$ and the vectors $\mathbf{b}^\mathcal{I}$,
$\mathbf{b}^\mathcal{E}$, $\mathbf{c}^\mathcal{E}$ and $\mathbf{c}^\mathcal{I}$
form the so-called Butcher tables of the IMEX Runge Kutta schemes given in
Appendix~\ref{time-scheme}.  Since the last lines of
$\mathbf{a}^{\mathcal{E},\mathcal{I}}$ are equal to
$\mathbf{b}^{\mathcal{E},\mathcal{I}}$ for PC2 and BPR353, the last operation
to retrieve $G_{\ell m}(r_k, t + \delta t)$ (Eq.~\ref{eq:assembly_imex}) is
actually redundant with the last sub-stage \citep[see][]{Ascher1997}.  IMEX
Runge-Kutta schemes require more computational operations to time advance the
set of equations(\ref{mass_conservation}--\ref{chemical_transport}) from $t$ to
$t+\delta t$ than CNAB2.  However, they allow larger time step sizes that
compensate for this extra numerical cost \citep{Marti2016} and they are more
accurate and stable.  Since BPR353 is a third-order scheme, it is particuraly
attractive to ensure an accurate equilibration of the most turbulent runs.

\subsection{Simulation diagnostics}
\label{sim_dia}
For each diagnostic quantity $f$, we adopt in the following overbars for time 
averaging and angle brackets for spatial averaging

\begin{equation*}
	\bar{f} = \dfrac{1}{t_\mathrm{avg}} \int_{t_0}^{t_0 + t_\mathrm{avg}} f(t) 
\,\mathrm{d}t \quad \mathrm{and} \quad \langle f \rangle_V = \dfrac{1}{V} 
\int_{V} f(\mathbf{r}, t) \,\mathrm{d}{V},
\end{equation*}
where $t_\mathrm{avg}$ corresponds to the averaging time.

\subsubsection{Integral quantities and scales}
The magnetic $E_m$ and kinetic $E_k$ energies are given by 

\begin{equation*}
	E_\mathrm{m}(t) + E_\mathrm{k}(t) = \dfrac{1}{2} 
\left[\displaystyle{\intoi} \dfrac{\magf^2(\mathbf{r}, t)}{EPm}\,\dV + 
	\displaystyle{\integralo} \vecu^2(\mathbf{r}, t)\,\dV \right],
\end{equation*}
where $\volumei$ is the inner core volume.
By multiplying the Navier-Stokes equation~(\ref{NS_dim}) by $\mathbf{u}$ and 
the induction equation~(\ref{induction}) by $\magf$, we obtain the 
following power balance
\begin{equation*}
	\frac{\mathrm{d}}{\mathrm{d}t} \left(E_\mathrm{k} + E_\mathrm{m}\right)(t) = \powerT(t) + \powerC(t) - \powerDiff(t) - \powerLor(t).
\label{power}
\end{equation*}
In the case of double-diffusive convection, the energy is provided by chemical 
and thermal buoyancy power $\powerC$ and $\powerT$ defined by 

\begin{equation*}
	\left\{
		\begin{array}{rp{0.01cm} lp{0.05cm}  lp{1cm}}
			\powerC(t) &=& V_o \left\langle \dfrac{\rac}{Sc} \xi(\mathbf{r}, t) \dfrac{r}{r_o}u_r(\mathbf{r}, t)\right\rangle_{V_o}\\[0.4cm]  
			\powerT(t) &=& V_o \left\langle\dfrac{\rat}{Pr} T(\mathbf{r}, t) \dfrac{r}{r_o}u_r(\mathbf{r}, t)\right\rangle_{V_o}\\[0.4cm]  
		\end{array}
	\right.
\end{equation*}
and dissipated by viscous and Ohmic dissipations $\powerDiff$ and $\powerLor$ 
given by

\begin{equation*}
	\left\{
		\begin{array}{r l  l}
			\powerDiff(t) &=& V_o\left\langle \left[ \grad \times \vecu(\mathbf{r}, t)\right]^2\right\rangle_{V_o}\\[0.4cm]
			\powerLor(t) &=& \left(V_o + V_i\right) \left\langle \dfrac{[\grad \times \magf(\mathbf{r}, t)]^2}{EPm^2}\right\rangle_{V_o + V_i}.
		\end{array}
	\right.
\end{equation*}

Once a statistically steady state has been reached, the input buoyancy powers
should compensate the Ohmic and viscous dissipations.  Following
\cite{King2012}, to assess the consistency of the numerical computations, we
measure the time-average difference between input and output powers $\Delta P$
\begin{equation*}
	\Delta P = 100 \times \frac{\overline{\powerT + \powerC - \powerDiff - \powerLor}}{\overline{\powerT + \powerC}}.
\end{equation*}
We made sure that this difference remained lower than $1.5\,\%$ for all the
simulations reported in this study. This value is below the $2~\%$ threshold
considered as sufficient to ensure the convergence of integral diagnostics
\citep[see][]{King2012,Gastine2015}.  For each simulation, the total convective
power $\powertot$ and the relative thermal convective power $\powerTrel$ are
defined by 

\begin{equation}
	\powertot = \overline{\powerT(t) + \powerC(t)} \  \mathrm{and} \  
\powerT^\% = \overline{{\dfrac{\powerT(t)}{\powerT(t) + \powerC(t)}}} \times 
100.
	\label{powert}
\end{equation}
$\powerTrel$ hence vanishes for a purely chemical forcing  and is 
equal to $100\,\%$ for a purely thermal forcing. 

Following \cite{Christensen2006} and \cite{Schwaiger2019}, we introduce two
quantities to characterise the typical flow lengthscale.  The integral scale
$L$ already discussed in the introduction is obtained from the time-averaged
kinetic energy spectrum

\begin{equation*}
	L = \pi \frac{2\overline{E_{\mathrm{k}}(t)}}{\displaystyle{\sum_{\ell} \ell \overline{\vecu^\ell(t) \cdot \vecu^\ell(t)}}},
\end{equation*}
where $\overline{\vecu^\ell \cdot \vecu^\ell}/2$ is the kinetic energy contained in spherical harmonic degree $\ell$, while the dominant lengthscale $\lpic$ is defined as the peak of the poloidal kinetic energy spectra \citep{Schwaiger2019,Schwaiger2020} 

\begin{equation*}
	\lpic = \mathrm{argmax}_\ell \left[ \overline{E_{k,p}^\ell(t)} \right],
\end{equation*}
where $E_{k,p}^\ell$ is the contribution of spherical harmonic degree $\ell$ to the total poloidal kinetic energy.

In order to explore the impact of the equatorial symmetry of the flow on the 
dipole-multipole transition, we consider the relative equatorially-symmetric 
kinetic energy $\ekrel$ introduced by \cite{Garcia2017}
\begin{equation}
	\label{ekrel}
	\ekrel = \overline{\dfrac{E_k^{\mathrm{s}, \mathrm{NZ}}}{E_k^\mathrm{NZ}}},
\end{equation}
where $E_k^\mathrm{NZ}$ is the kinetic energy contained in the non-zonal flow 
and $E_k^{\mathrm{s},\mathrm{NZ}}$ the kinetic energy contained in the 
equatorially-symmetric part of the non-zonal flow.

\par We measure the mean convective flow amplitude either by the magnetic 
Reynolds number $Rm$ or by the Rossby number $Ro$ defined by 
\begin{equation*}
	Rm = \overline{\sqrt{2E_k(t)}}Pm = \dfrac{\tau_\eta}{\tau_\mathrm{adv}},\ Ro = \overline{\sqrt{2E_k(t)}}E= \dfrac{\tau_\Omega}{\tau_\mathrm{adv}},
\end{equation*}
where $\tau_\mathrm{adv} = \lscale/u_{\mathrm{rms}}$ is the characteristic turnover time and 
$u_{\mathrm{rms}}$ the root mean square  flow velocity. 
Following \cite{Christensen2006}, we define a local Rossby number by 
\begin{equation*}
	Ro_L =  Ro\frac{\lscale}{L}.
\end{equation*}

\par The dipolar character of the CMB magnetic field is quantified by its
dipolar fraction $f_\mathrm{dip}$, defined as the ratio of the axisymmetric
dipole component to the total field strength at the CMB in spherical harmonics
up to degree and order $12$.

\begin{figure}
	\centering
	\includegraphics[width=\linewidth]{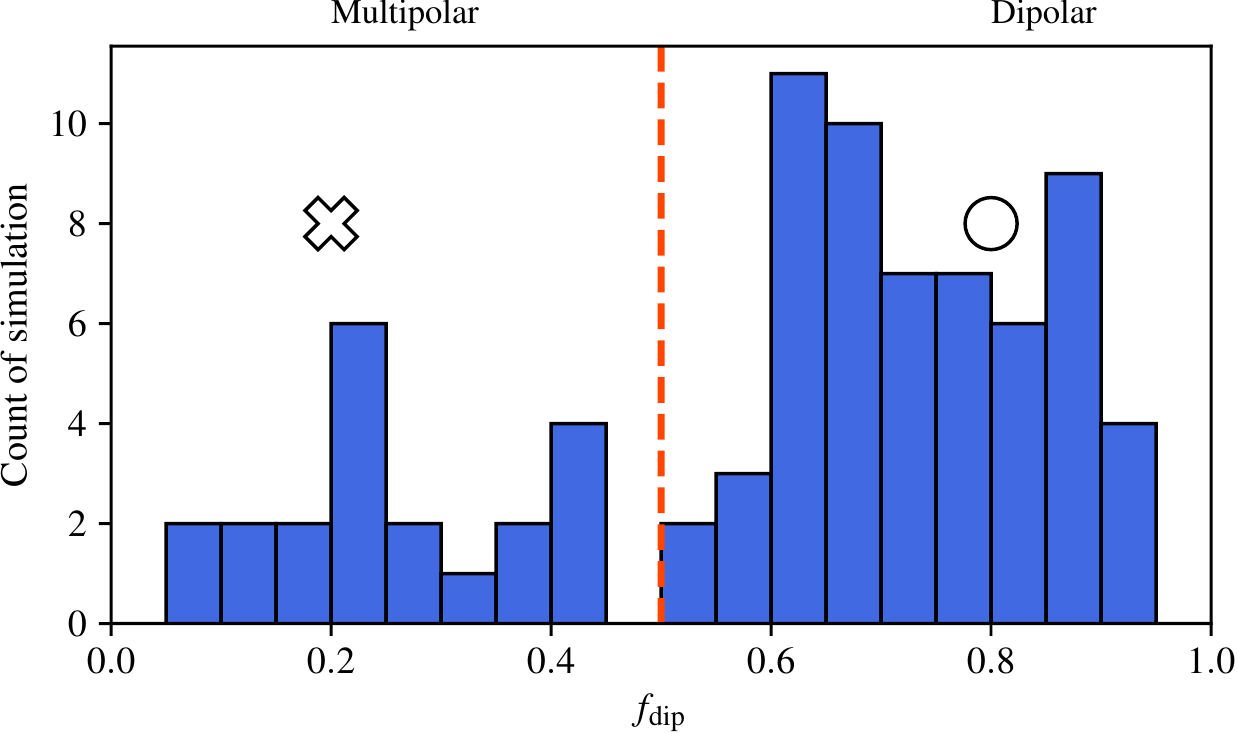}
	\caption{Distribution of the dipolar fraction $\fdip$ for the 
		$79$ numerical simulations of this study. Multipolar simulations 
		are defined as having  $\fdip < 0.5$, and they will be marked 
		by a cross in subsequent figures. Dipolar simulations ones will 
		be displayed using a circle. The vertical dashed line marks 
		the $\fdip=0.5$ limit between dipolar and multipolar 
		simulations.}
	\label{histo}
\end{figure}

Figure~\ref{histo} shows the statistical distribution of $\fdip$ for the
simulations reported in this study.  Two distinct groups of numerical
simulations separated by a gap at $\fdip \approx 0.5$ are visible in this
figure.  A magnetic field is considered as dipolar when $\fdip > 0.5$ and
multipolar otherwise.  This bound differs from the original threshold of $\fdip
= 0.35$ considered by \cite{Christensen2006}, but it is found to better
separate the two types of dynamo models contained in our dataset. Note that the
same bound of 0.5 was recently chosen by \cite{Menu2020} in their study.  The
magnetic field amplitude is measured by the Elsasser number $\Lambda$ 
\begin{equation*}
	\Lambda = \overline{\sqrt{2E_m}}.
\end{equation*}

Finally, transports of heat and chemical composition are quantified by using
the Nusselt $Nu$ and the Sherwood $Sh$ number defined by
\citep[see][chapter~1]{Goluskin2016}

\begin{equation*}
	Nu = \frac{\Delta T_0}{\overline{\Delta T}} \quad \mathrm{and} \quad Sh = \frac{\Delta \xi_0}{\overline{\Delta \xi}},
\end{equation*}
where $\Delta T$ , $\Delta \xi$ are the temperature and composition differences
between the ICB and the CMB, and the subscript $0$ stands for the background
conducting state.  For both the fixed fluxes and hybrid configurations (recall
section~\ref{sec:bc} above), we obtain a background composition contrast given
by 
\begin{equation}
 \Delta \xi_{0} = \frac{\eta(\eta + 2)}{2(\eta^2 + \eta + 1)},
\end{equation}
where $\eta=r_i/r_o$ is the radius ratio.  The background temperature drop
$\Delta T_0$ depends on the imposed boundary conditions. For the fixed flux
setup, it reads
\begin{equation}
	\label{delta_T1}
	\Delta T^\mathrm{ff}_0 =  \frac{\eta(\eta + 2)}{2(\eta^2 + \eta + 1)}=\Delta \xi_0,
\end{equation}
while it becomes
\begin{equation}
	\label{delta_T2}
	\Delta T^\mathrm{hyb}_0 = \frac{1}{\eta},
\end{equation}
for the hybrid setup.

Table~\ref{output} gives the definition of most of these integral diagnostics
and provides estimates for Earth's core, along with the bracket of values
obtained in the numerical dataset presented here. 

\subsection{Exploration of parameter space}
\label{sec:explo}
We compute $79$ simulations varying the Ekman number, the magnetic Prandtl
number, the thermal and the chemical Rayleigh numbers. The properties of this
dataset are listed in  Tab.~\ref{simu_tab}.  The less turbulent simulations
have been initialized with a strong dipolar field and a random thermo-chemical
perturbation.  Their final states have been used as initial conditions for the
more turbulent simulations, in order to shorten their transients.  Three
different Ekman numbers are considered in this study : $10^{-4}$, $3 \times
10^{-4}$ and $10^{-5}$.  $\rat$ has been varied between $0$ and $6 \times
10^{10}$ and $\rac$ between $0$ and $1.9 \times 10^{12}$ to study the influence
of the convective forcing and span the transition between dipole-dominated and
multipolar dynamos.  We adopt $Pr = 0.3$ and $Sc = 3$ (i.e. $Le=10$) for a
better comparison with previous studies \citep[e.g.][]{Takahashi2014} and to
mitigate the computational cost associated with large Lewis numbers. $Pm$
varies between $0.5$ and $7$, depending on the Ekman number, in order to
maintain $Rm > 100$.  The numerical models were integrated for at least 20~\%
of a magnetic diffusion time $\tau_\eta$ for the most turbulent (and demanding)
ones, and  for more than one $\tau_\eta$ for the others, in order to ensure
that a statistically steady state had been reached.

%% file: seuil.tex
\section{Results}
\label{results}
\subsection{Onset of top-heavy convection}
\label{onset}
\begin{figure*}
\centering
\includegraphics[width=\linewidth]{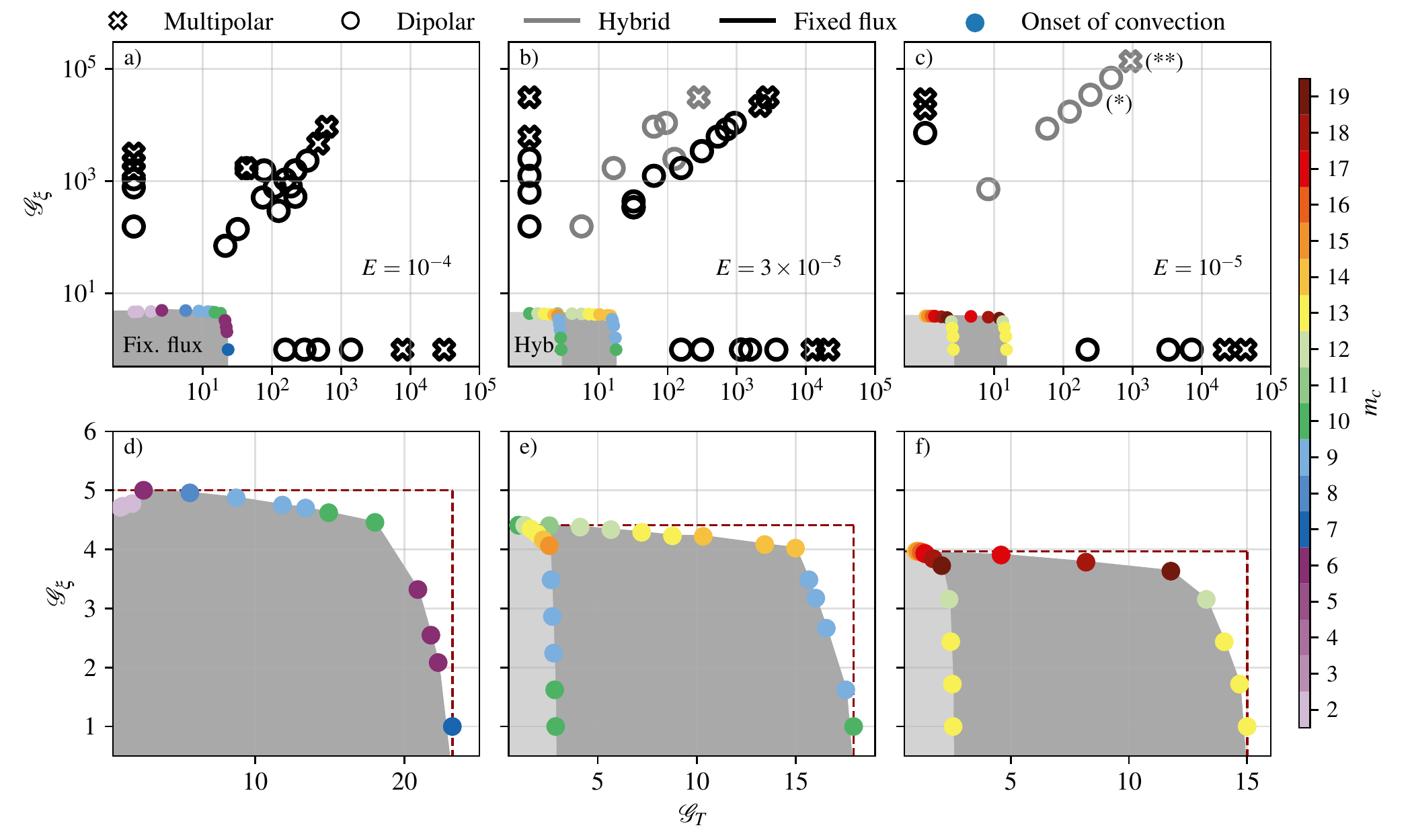}
\caption{Linear onset of top-heavy convection in 
	($1 + Gr_T E^{4/3}$, $1 + Gr_\xi E^{4/3}$) parameter
	space, where $Gr_T$, $Gr_\xi$, and $E$ are the thermal Grasshof, 
	chemical Grasshof and Ekman numbers, respectively, 
	for the three Ekman numbers $E$ considered in this study: 
	$10^{-4}$ (left column), $3\times10^{-5}$ (center column), 
	and $10^{-5}$ (right column). Critical curves correspond to the edges 
	of the gray shaded areas. Dark gray areas were obtained for fixed-flux 
	boundary conditions and light gray areas for hybrid boundary conditions, 
	the latter present only for $E = 3\times10^{-5}$ and $E=10^{-5}$. 
	The bottom panels (d), (e) and (f) show zoomed-in insets of upper panels 
	(a), (b) and (c). The edges of the gray areas, which define the critical 
	curves, connect discs whose color defines the critical azimuthal wavenumber 
	$m_c$. The top row (in logarithmic scale in both directions) also features
	the location in parameter space of the $79$ simulations computed in this 
	study. Circles (resp. crosses) represent dipolar (resp. multipolar) 
	simulations. Circles and crosses with gray (resp. black) edges correspond 
	to fixed-flux (resp. hybrid) boundary conditions. Simulations (*) and 
	(**) are reference simulations discussed in detail in the text
	(see also Tab.~\ref{simu_tab}). 
}
\label{ra_raxi}
\end{figure*}

Figure~\ref{ra_raxi} shows the location of the $79$~computed numerical 
simulations for the three considered Ekman numbers in the parameter space 
($\mathcal{G}_T$, $\mathcal{G}_\xi$) defined by 

\begin{equation*}
	\mathcal{G}_T = 1 + Gr_T E^{4/3} \quad \mathrm{and} \quad \mathcal{G}_\xi = 1 + Gr_\xi E^{4/3},
\end{equation*}
where $Gr_{T(\xi)}$ corresponds to the thermal (chemical) Grasshof number, 
\begin{equation*}
	Gr_{T} = \dfrac{\rat}{Pr} \quad \mathrm{and} \quad Gr_{\xi} = \dfrac{\rac}{Sc}.
\end{equation*}
Adding $1$ to $Gr_{T (\xi)} E^{4/3}$ in the above equations allows us to use 
logarithmic scales in the top row of Fig.~\ref{ra_raxi}.

We determine the onset of convection using the open-source software
SINGE\footnote{\url{https://bitbucket.org/vidalje/singe}} which computes linear
eigenmodes for incompressible, double-diffusive fluids enclosed in a spherical
cavity \citep[see][]{Schaeffer2013, Vidal2015, Kaplan2017, Monville2019}.  For
a fixed $\rat$ ($\rac$), the code solves the generalized eigenvalue problem
formed by the linearized Navier-Stokes and transport equations.  It seeks
eigenmodes $f$ of the form
\begin{equation*}
	f(t, r, \theta, \phi) = \hat{f}(r, \theta)\exp[i(m\phi - \omega t)],
\end{equation*}
where $\hat{f}$ is a function of $r$ and $\theta$, $m$ is the azimuthal wave
number and $\omega$ the complex angular frequency.  Starting at a specific
($\rat$, $\rac$), the critical mode is determined by varying one of the
Rayleigh numbers (keeping the other fixed) in order to obtain an $\omega$ with
a vanishing imaginary part.  The onset mode is then characterized by its
critical chemical and thermal Rayleigh numbers ($\rac^c$, $\rat^c$), its
critical azimuthal wave number $m_c$ and its (real) angular drift frequency
$\omega_c$.  Connecting the ($\rac^c$, $\rat^c$) pairs  that we collected with
SINGE gives rise to the critical curves plotted in Fig.~\ref{ra_raxi}.  In each
panel, the intersection of these  curves with the $x$-axis (resp. $y$-axis)
corresponds to the critical Rayleigh number for purely thermal (resp. chemical)
convection.  Underneath the critical curves, the grey-shaded areas are regions
of the parameter space where thermal and chemical perturbations are unable to
trigger a convective flow.

We note that the shape delimited by the critical curves in the bottom row of
Fig.~\ref{ra_raxi} is not rectangular, as would be the case if the two sources
of buoyancy were independent.  Instead, we observe a decrease of the critical
$\mathcal{G}^c_\xi$ when $\mathcal{G}^c_T$ increases for the three different
Ekman numbers.  As previously reported by \cite{Busse2002}, \cite{Trumper2012}
and \cite{Net2012}, and discussed in the Introduction, this demonstrates that
the addition of a second buoyancy source facilitates the onset of convection as
compared to the single diffusive configurations. 

Starting from $\mathcal{G}^c_T = 0$ and following the critical curve for each
Ekman number, $m_c$ grows until one reaches the upper right "corner'' of the
onset region and then decreases to a value comparable to the starting
$\mathcal{G}^c_T = 0$ $m_c$ value when $\mathcal{G}^c_\xi$ tends to zero.  We
observe that $m_c$ is nearly constant on the vertical branches, while it
increases much faster on the horizontal ones, as already reported by
\cite{Trumper2012}. 

As can be seen in Figs.~\ref{ra_raxi}(e) and~\ref{ra_raxi}(f), the shaded shape
is much wider with fixed-flux boundary conditions than with hybrid boundary
conditions. This comes from the difference in the temperature contrasts of the
background conducting states.  An adequate way to compare both setups resorts
to using \emph{diagnostic} Grasshof numbers
\begin{equation}
	\widetilde{Gr_T} = Gr_T \Delta T_0,\quad
	\widetilde{Gr_\xi} = Gr_\xi \Delta \xi_0, 
\end{equation}
that are based on the temperature (composition) contrasts  $\Delta T_0$ ($
\Delta \xi_0$) of the reference conducting state instead of the \emph{control}
thermal (chemical) Grasshof numbers $Gr_T$  ($Gr_\xi$)
\citep[see][]{Johnston2009,Goluskin2016}.  Using Eqs.~(\ref{delta_T1}) and
(\ref{delta_T2}), the temperature difference between ICB and CMB for the
conducting states for both setups reads
\begin{equation*}
	\frac{\Delta T_0^\mathrm{hyb}}{\Delta T_0^\mathrm{ff}} = \frac{2 (\gamma^2 + 
\gamma + 1)}{\gamma^2 (\gamma + 1)} \approx 10.23\,,
\end{equation*}
which is in good agreement with the actual ratio of critical $\mathcal{G}^c_T$ 
observed in panels (e) and (f) of Fig.~\ref{ra_raxi}.

\begin{figure}
	\centering
	\includegraphics[width=\linewidth]{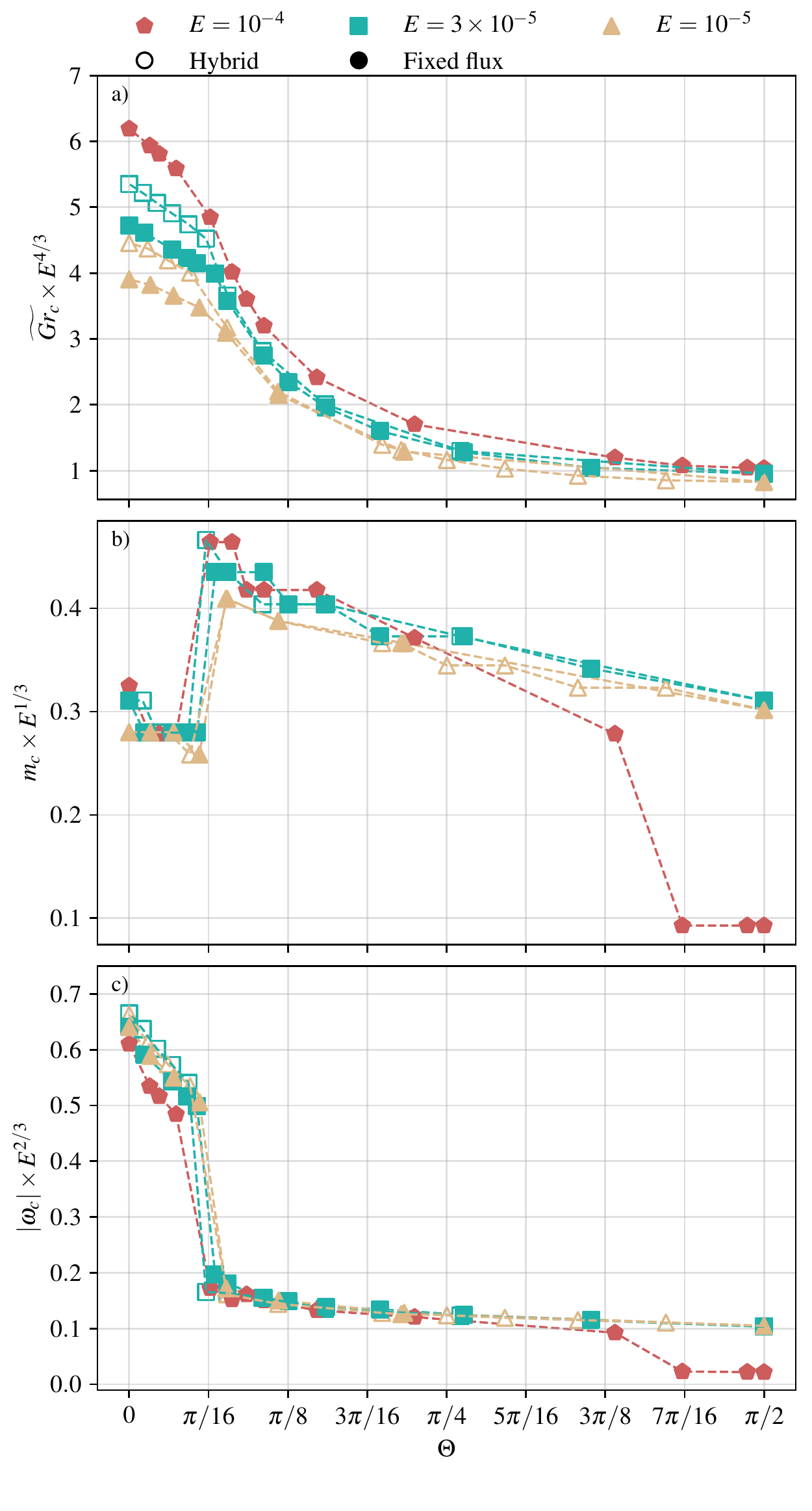}
	\caption{(a) Critical effective Grasshof number multiplied by $E^{4/3}$ as 
a function of the Grasshof mixing angle $\Theta$. (b) Critical azimuthal 
wavenumber $m_c$ multiplied by $E^{1/3}$ as a function of $\Theta$. (c) 
Critical drift frequency $\omega _c$ multiplied by $E^{2/3}$ as a function of 
$\Theta$. Symbol colors correspond to the Ekman number, open-symbols to hybrid 
boundary conditions and filled symbols to fixed-flux boundary conditions.}
	\label{silva}
\end{figure}

The analysis of the onset of double-diffusive convection becomes even more
straightforward if one adopts the formalism introduced by \cite{Silva2019}.
This framework rests on two parameters: first, the \emph{diagnostic effective
Grasshof number}

\begin{equation}
	\widetilde{Gr_c} = \sqrt{\left( \widetilde{Gr^c_T} \right) ^2 + 
	\left( \widetilde{Gr^c_\xi} \right) ^2},
\end{equation}
and second the \emph{Grasshof mixing angle} $\Theta$ such that
\begin{equation}
\cos\Theta = \dfrac{\widetilde{Gr^c_T}}{\widetilde{Gr_c}}, \quad
\sin\Theta = \dfrac{\widetilde{Gr^c_\xi}}{\widetilde{Gr_c}}.
\end{equation}
The pair $(\widetilde{Gr_c},\Theta)$ can be interpreted as the polar
coordinates of the critical onset mode in the
$(\widetilde{Gr_T},\widetilde{Gr_\xi})$ Cartesian parameter space. A mixing
angle $\Theta=0$ hence corresponds to purely thermal convection, while
$\Theta=\pi/2$ corresponds to purely chemical convection.  Figure~\ref{silva}
shows the critical effective Grasshof number $\widetilde{Gr_c}$, the critical
azimuthal wave number $m_c$ and the critical angular drift frequency
$|\omega_c|$, multiplied in each instance by their expected asymptotic
dependence on the Ekman number for purely thermal convection, as a function of
the Grasshof mixing angle  $\Theta$.  Adopting a diagnostic effective Grasshof
number conveniently enables the merging of the onset curves associated with the
two sets of thermal boundary conditions (fixed-flux and hybrid).

The onset curves can be separated into two branches: (\textit{i}) From $\Theta
=0$ up to $\Theta \approx \pi/16$, the onset mode almost behaves as a pure
low-$Pr$ thermal mode with little change in $m_c\, E^{1/3}$ and a large drift
speed; (\textit{ii}) a sharp transition  to another kind of onset mode,
reminiscent of the $Pr\gtrsim 1$ convection onset, is observed for $\Theta
\gtrsim \pi/16$. The latter is characterized by  a smaller drift frequency, a
higher azimuthal wavenumber and a lower effective Grasshof number
$\widetilde{Gr_c}$. The critical azimuthal wavenumber reaches its maximum for
$\Theta \approx \pi/16$ before gradually decreasing to reach a value comparable
to that expected for purely thermal convection towards $\Theta =\pi/2$.

The Ekman dependence is almost perfectly captured by the asymptotic scalings
$E^{-1/3}$ and $E^{-2/3}$ for the critical wavenumber $m_c$ and the drift
frequency $\omega_c$. The case of mixing angles $\Theta \geq 7\pi/16$ with
$E=10^{-4}$ constitutes an exception to this rule with a sharp drop to a
constant critical wavenumber $m_c=2$ (see the small kink in the upper left part
of Fig.~\ref{ra_raxi}d).  The dependence of $\widetilde{Gr_c}$ on the Ekman
number shows a more pronounced departure from the leading-order asymptotic
scaling $Gr_c \sim E^{-4/3}$. As shown by \cite{Dormy2004} for differential
heating (see e.g. their Fig.~5), the higher-order terms in the asymptotic
expansion of $Ra_T^c$ as a function of the Ekman number are still significant
for $E>10^{-6}$ \citep[see also][]{Schaeffer2016fgsh}.  Given the range of
Ekman numbers considered for this study, it is not suprising that the
asymptotic scaling for $Gr_c$ is not perfectly realized yet. 

In summary, the onset of double-diffusive convection in the top-heavy regime
takes the form of thermal-like drifting Rossby-waves, the nature of which
strongly depends on the fraction between chemical and thermal forcings. This
confirms the results previously obtained by \cite{Silva2019} (their Fig.~9).

%% file: earth_like.tex
\subsection{A reference case}
\label{ref_case}
\begin{figure*}
	\centering
	\includegraphics[width=\linewidth]{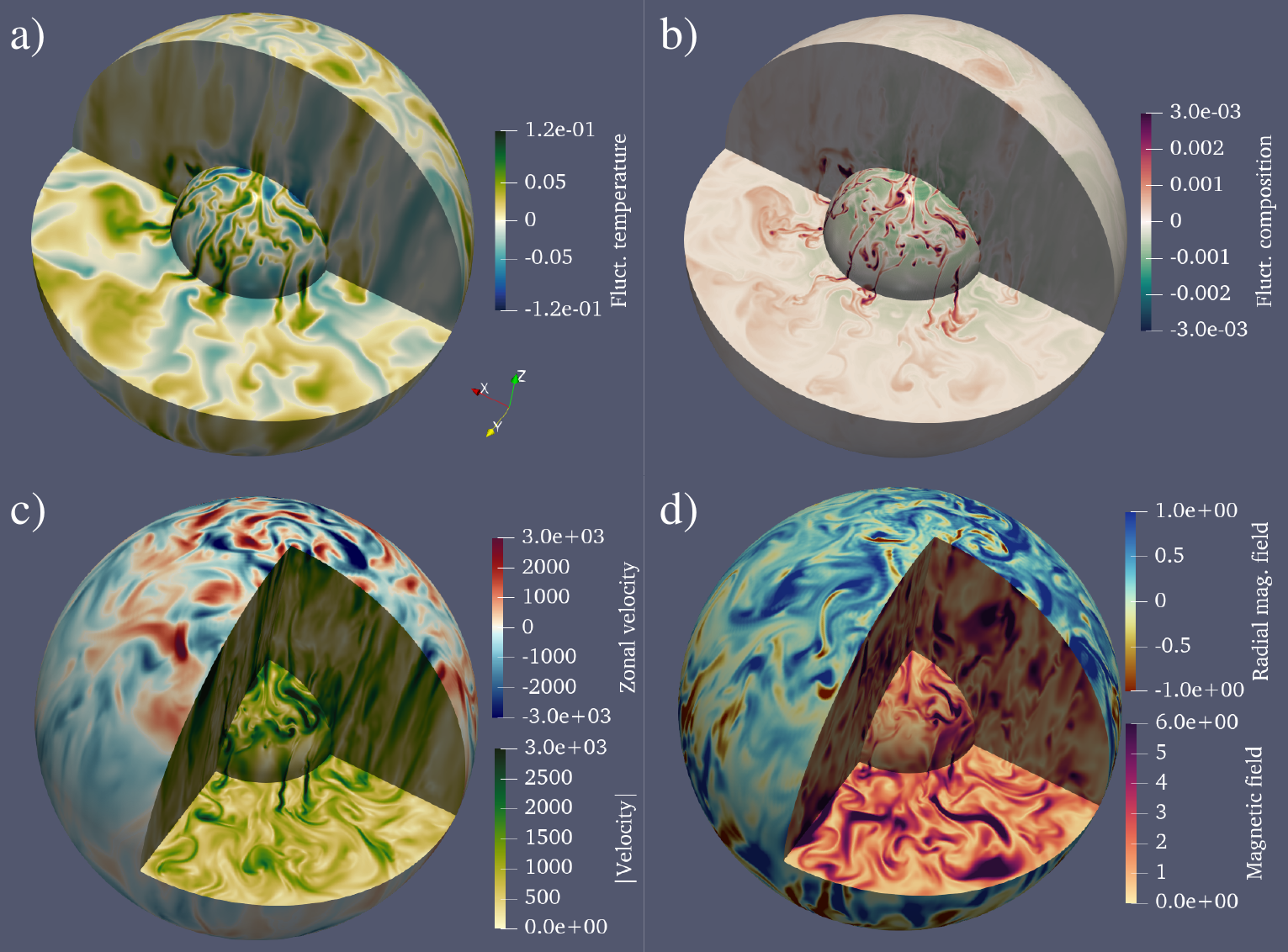}
	\caption{Three-dimensional renderings of a snapshot of simulation (*) 
		(see Tab.~\ref{simu_tab}). On each panel, the inner and outer 
		spherical surfaces correspond to dimensionless radii 
		$r = 0.57$ and $r = 1.5$, respectively. The $z$ axis displayed 
		in panel (a) corresponds to the axis of rotation. (a): Temperature 
		perturbation. (b): Composition. (c): Velocity. The outer surface 
		shows the zonal velocity $u_\varphi$, while the inner surface, 
		the equatorial cut and the two meridional cuts display the magnitude
		of the velocity field, $|\vecu|$. (d): Magnetic field. 
		The outer surface shows the radial field $B_r$, while the inner 
		surface, the equatorial cut and the two meridional cuts display 
		the magnitude of the magnetic field, $|\magf|$. All fields 
		are dimensionless.}
	\label{reference}
	
	\centering
	\includegraphics[width=\linewidth]{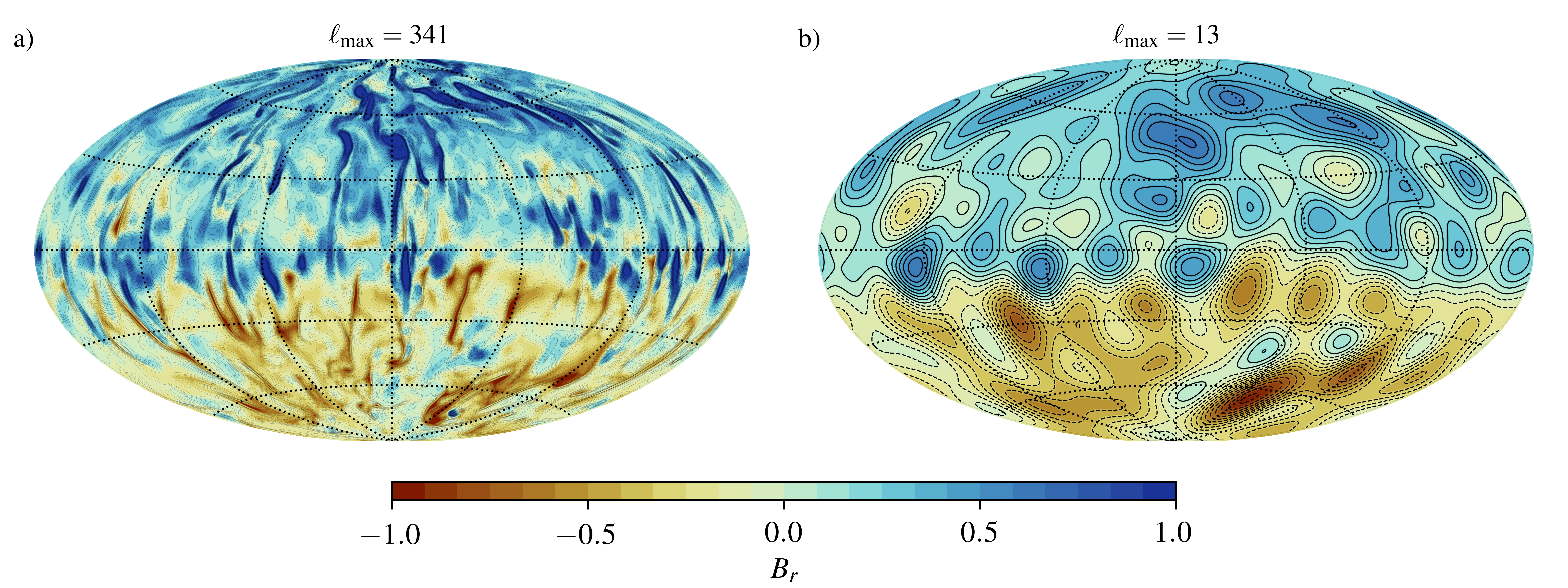}
	\caption{Hammer projection of the dimensionless radial magnetic field at the 
		CMB truncated at the spherical harmonic degree $\ell_\mathrm{max} = 341$ 
		(a) and at spherical harmonic degree $\ell_\mathrm{max} = 13$ 
		(b) for the numerical simulation (*) shown in Figure~\ref{reference}. 
		It corresponds to the same snapshot as in Figure~\ref{reference}.
		The dashed lines in panel(b) correspond to negative values of 
		the radial magnetic field.}
	\label{B_surf}
\end{figure*}

We will now focus on the simulation marked by an asterisk (*) in
Table~\ref{simu_tab} and in Figure~\ref{ra_raxi} to highlight specific
double-diffusive convection features.  This simulation corresponds to $\rat =
3.4 \times 10^8$, $\rac = 4.8 \times 10^{11}$, $E = 10^{-5}$ and $Pm = 0.5$
with hybrid boundary conditions.  The thermal convective power amounts on
average for $46\,\%$ of the total input power.  Since the local Rossby number
$Ro_L$ reaches $0.11$, this simulation is expected to operate in a parameter
regime close to the transition between dipolar and multipolar regimes ($Ro_L
\approx 0.12$) put forward by \cite{Christensen2006}.  This high value of
$Ro_L$ also indicates the sizeable role played by inertia in the force balance
of this simulation.  Figure~\ref{reference} shows 3-D renderings of several
fields extracted from a snapshot taken over the course of the numerical
integration: (a) temperature perturbation, (b) composition, (c) zonal velocity
and magnitude of the velocity vector, (d) radial magnetic field and magnitude
of the magnetic field vector. We chose the radius of the inner and outer
spheres of these renderings to place ourselves outside thermal and
compositional boundary layers.  Convection is primarily driven by the chemical
composition flux at the ICB.  Because of the contrast in diffusion coefficients
($Le=10$), compositional plumes develop at a much smaller scale than that of
thermal plumes (see Fig.~\ref{reference}a and Fig.~\ref{reference}b). Having
$Le=10$ also induces a chemical boundary layer much thinner than the thermal
one, as illustrated  by the Sherwood number $Sh$ being five times larger than
the Nusselt number $Nu$  in this case (44.8 versus 8.0, see
Tab.~\ref{simu_tab}).  The emission of a thermal plume is likely triggered by
an impinging chemical plume. Accordingly, one would expect temperature
fluctuations to be enslaved to compositional fluctuations, which may explain
the outstanding spatial correlation between temperature and composition
noticeable in Fig.~\ref{reference}(a) and Fig.~\ref{reference}(b).  This
correlation was already reported by \cite{Trumper2012} for predominantly
thermal convection. 

In typical dipole-dominated dynamos, the velocity field features extended
sheet-like structures that span most of the core volume
\citep[see][]{Yadav2013, Schwaiger2019}.  Because of the strong forcing that
characterizes the reference simulation, the convective flow is organised on
smaller scales, which likely reflects  the sizeable amplitude of inertia.  The
magnetic field is dominated by its axisymmetric dipolar component (see
Fig.~\ref{B_surf}a), despite the supposed proximity of the simulation with the
dipolar-multipolar transition zone.  Inspection of Table~\ref{simu_tab}
indicates that this snapshot-based observation can in fact be extended to the
entire duration of the simulation (close to half a magnetic diffusion time), as
the average dipolar fraction $\fdip$ is equal to $0.77$. 

The radial magnetic field features intense localized flux patches of mostly
normal polarity in each hemisphere, with a few reverse flux patches paired with
normal ones, mostly at high latitudes.  Those fluid regions hosting a locally
strong magnetic field are characterized by more quiescent flows, as can be seen
in the equatorial planes of Fig.~\ref{reference}(c) and Fig.~\ref{reference}(d)
in the vicinity of the outer boundary.  In a global sense, our reference
simulation can be qualified as a {\em strong-field} dynamo since the ratio of
the magnetic energy $E_m$ to kinetic energy $E_k$ is slightly larger than
unity, $E_m/E_k=1.36$. 

Figure~\ref{B_surf} shows a comparison between the radial component of the
magnetic field at the CMB (left) and its representation  truncated at spherical
harmonic degree  $\ell_\mathrm{max} = 13$ (right).  The truncated field
presents several key similarities with the geomagnetic field at the CMB, such
as a significant axial dipole and patches of reverse polarity in both
hemispheres. 

\subsection{Earth-likeness}
\label{chi}
\begin{figure*}
	\centering
	\includegraphics[width=\linewidth]{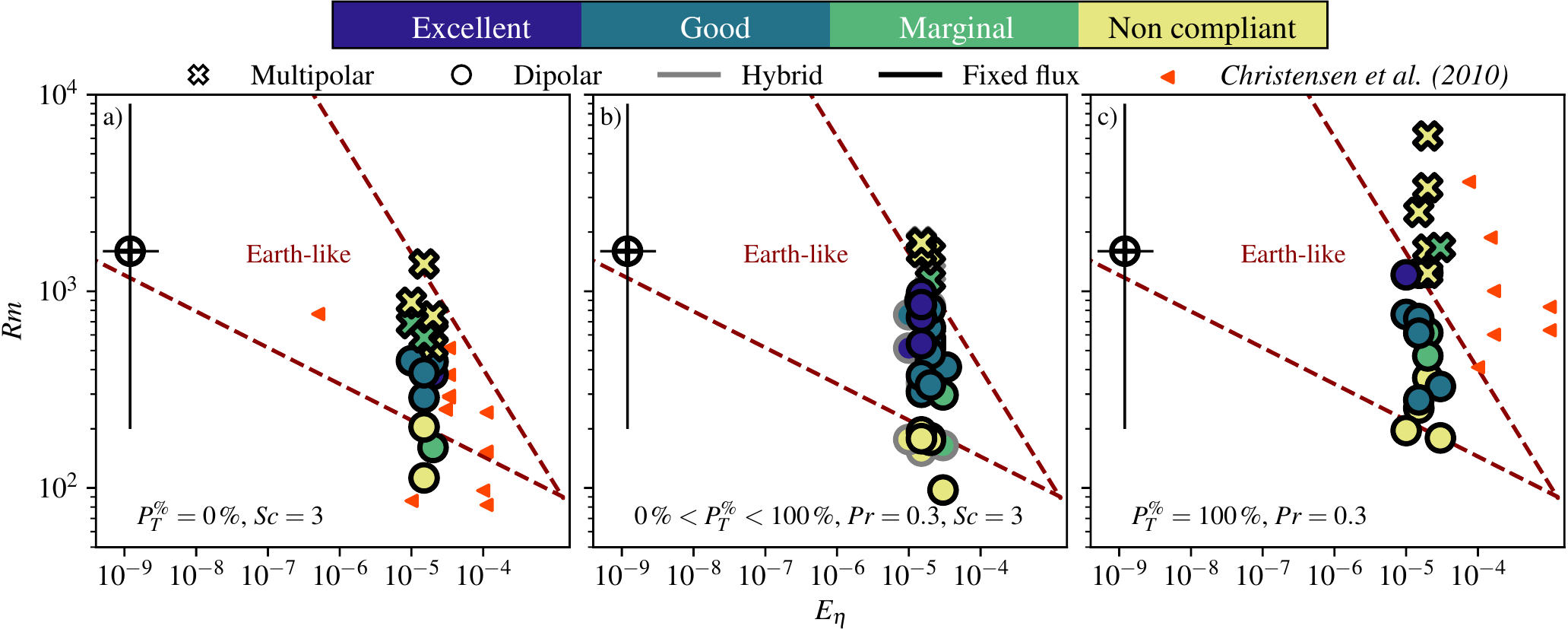}
	\caption{Compliance parameter $\chi^2$ as a function 
		of the magnetic Ekman number $E_\eta$ and of the magnetic Rayleigh 
		number $Rm$ for three different setups: purely chemical 
		convection (left), double-diffusive convection (center) and 
		pure thermal convection (right). 
		Dashed lines mark the limits of the 
		Earth-like domain defined by 
		\cite{Christensen2010} and the black symbol $\bigoplus$ 
		marks the approximate position of the Earth's dynamo in this 
		representation. The significant size of the error bars is due 
		to the wide range of estimates for $u_{\mathrm{rms}}$  and 
		$\eta$ (see Tab.~\ref{dimensionless}). The
		triangles correspond to the simulations provided by 
		\cite{Christensen2010} with $Pr = 0.3$ or $Sc = 3$. 
}
	\label{wedge}

\end{figure*}

For a more quantitative assessment of the Earth-likeness of the dynamo models,
we employ the rating of compliance $\chi^2$ introduced by
\cite{Christensen2010}.  This quantity is derived from four criteria based on
the magnetic field at the CMB truncated at the degree $\ell=8$

\begin{enumerate}
	\item the relative axial dipole energy $AD/NAD$, which corresponds to the 
ratio of the magnetic energy in the axial dipole field to that of the rest of 
the field up to degree and order eight,
	\item the equatorial symmetry $O/E$ corresponding to the ratio of the 
magnetic energy at the CMB of components that have odd values of $(\ell + m)$ 
for harmonic degrees between two and eight to its counterpart in components 
with $\ell + m$ even,
	\item the zonality $Z/NZ$, which corresponds to the ratio of the zonal to 
non-zonal magnetic energy for harmonic degrees two to eight at the CMB,
	\item the flux concentration factor $FCF$, defined by the
variance in the squared radial field.
\end{enumerate}
To evaluate these quantities for the Earth, \cite{Christensen2010} used
different models based on direct measures such as \textit{gufm1} model by
\cite{Jackson2000} and IGRF-11 model \citep[from][]{Finlay2010a}, as well as
archeomagnetic and lake sediment data \citep[model CALS7K.2 from][]{Korte2005}
and a statistical model for paleofield \citep[see][]{Quidelleur1996}.  These
models allow to estimate the evolution of the mean value of the Gauss
coefficients and their variances.  Finally, they obtained the values given in
Table~\ref{criteria} for the four rating parameters.
\begin{table}
\centering
\caption{Time-average of the rating parameters defined by \cite{Christensen2010} 
	 for Earth and the simulation (*) (see Tab.~\ref{simu_tab}).}
\label{criteria}
\begin{tabular}{c c c}\hline
	\textbf{Name} &       \textbf{Earth's value} &   \textbf{Simulation (*)} \\ \hhline{===}
       $AD/NAD$      &1.4&2.69\\
       $O/E$   &1.0&1.75\\
       $Z/NZ$  &0.15&0.25\\
       $FCF$   &1.5&1.46\\
\end{tabular}
\end{table}
These values are used to determine the rating of compliance between numerical 
dynamo models and the geomagnetic field $\chi^2$ expressed by
\begin{equation*}
	\chi^2 = \sum_{\psi_k} \left\{ \frac{\ln(\overline{\psi_k}) - \ln(\psi_k^{\earth})}{\ln\left[\sqrt{\mathrm{Var}\left(\psi_k^{\earth} \right)} \,\right]} \right\} ^2,
\end{equation*}
where $\psi_k \in \{AD/NAD, O/E, Z/NZ, FCF\}$, Var($\psi_k^{\earth}$) is the
variance of $\psi_k^{\earth}$ and the exponent $\earth$ stands for the Earth
core.  The agreement between simulation and Earth is termed by
\cite{Christensen2010} as \emph{excellent} if $\chi^2 <2$, as \emph{good} when
$2 \le \chi^2 \le 4$, as \emph{marginal} wen $4 < \chi^2 \le 8$ and
\emph{non-compliant} when $\chi^2 > 8$.  We adopt the same classification in
the following.  According to Table~\ref{criteria}, the relative axial dipole
power ($AD/NAD$) and the equatorial symmetry ($O/E$) of the reference
simulation (*) are too large in comparison with the reference geomagnetic
values, which penalises the overall compliance of the simulation.
Nevertheless, the simulation remains in excellent agreement with the
geomagnetic field with $\chi^2 = 1.5$.

Based on this rating of compliance $\chi^2$, \cite{Christensen2010} propose a
representation to classify the different numerical dynamos according to the
ratio of three different timescales: the rotation period $\tau_\Omega$, the
advection time $\tau_\mathrm{adv}$ and the magnetic diffusion time $\tau_\eta$.
Those can be cast into two dimensionless numbers: the magnetic Reynolds number
$Rm$ defined by $\tau_\eta/\tau_\mathrm{adv}$ and the magnetic Ekman number
$E_\eta$  defined by $\tau_\Omega/\tau_\eta = E/Pm$.  Figure~\ref{wedge} shows
our numerical simulations plotted in the parameter space ($E_\eta$,$Rm$).  For
comparison purposes, the simulations by \cite{Christensen2010} with $Pr = 0.3$
or $Sc = 3$ (triangular markers) have been added to our $79$ dynamo models.  To
single out the effect of the diffusivities, the purely chemical, the
double-diffusive and the purely thermal simulations have been plotted
separately.  The black symbol marks the approximate position of Earth's dynamo
in this representation (see Tab.~\ref{phy_param} and Tab.~\ref{output}).
\cite{Christensen2010} posited the existence, in this parameter space, of a
triangular wedge (delimited by dashed lines in Fig.~\ref{wedge}), inside which
the numerical dynamos yield Earth-like surface magnetic fields (those from our
dataset are shown in Fig.~\ref{wedge} with dark blue and blue disks for
Excellent and Good ratings, respectively).  Below the wedge, weakly
super-critical dynamos feature too dipolar magnetic  fields,  on the account of
the modest value of the  magnetic Reynolds number.  Conversely, a significant
increase of the input power at a given $E_\eta$ yields small-scale convective
flows, which possibly lead to the breakdown of the axial dipole (multipolar
simulations displayed with crosses in Fig.~\ref{wedge}).  We observe in
Fig.~\ref{wedge} that the upper boundary of the wedge actually depends on the
value of $\powerTrel$: while several purely chemical simulations are already
multipolar below $Rm \simeq 500$, the transition to multipolar dynamos is
delayed to $Rm > 1500$ for the purely thermal ones.  Although all those dynamo
models that possess a good or  excellent semblance with the geomagnetic field
at the CMB lie within the boundaries of the original wedge, having a pair
$(E_\eta,Rm)$ that lies within this wedge cannot be considered as a sufficient
condition to produce an Earth-like magnetic field, since the wedge includes a
number of simulations with marginal or poor semblance.

\begin{figure*}
	\centering
	\includegraphics[width=\linewidth]{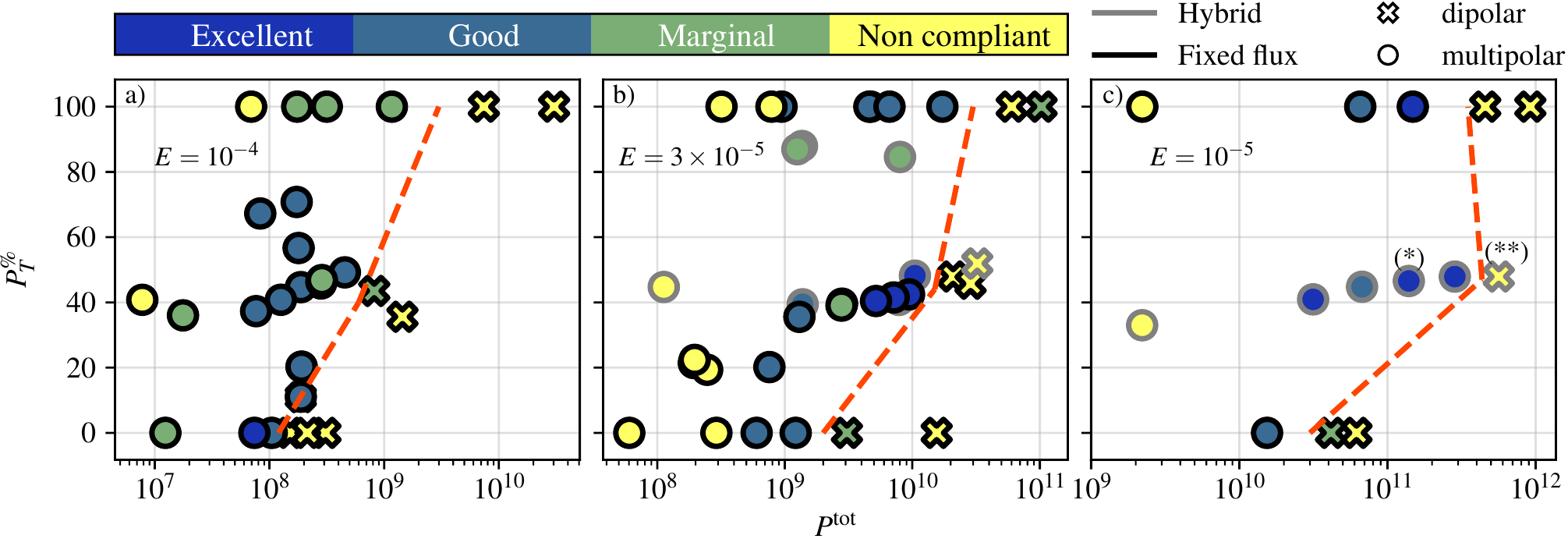}
	\caption{Compliance parameter $\chi^2$ as a function of the input power
		$\powertot$ and of the relative thermal buoyancy power
		$\powerT^\%$ for the three different Ekman numbers considered 
		in this study. 
		The different colors correspond to the compliance quality defined 
		by \cite{Christensen2010}.
		The dashed lines mark a tentative extrapolation of the 
		transition between dipolar and multipolar dynamos in this parameter space.}
	\label{pt_ptot}
\end{figure*}
To further discuss the impact of $\powerT^\%$ on the morphology of the magnetic
field at the CMB, Fig.~\ref{pt_ptot} shows $\chi^2$ in the parameter space
($\powertot$, $\powerT^\%$) for the three different Ekman numbers considered
here.  The dashed lines mark the tentative boundaries between dipolar and
multipolar simulations in term of $\powertot$.  The dipole-multipole transition
is delayed to larger input power $\powertot$ at lower Ekman numbers.
Decreasing $E$ indeed enables the exploration of a physical regime with lower
$Ro$ prone to sustain dipole-dominated dynamos
\citep[see][]{Kutzner2002,Christensen2006}. The input power required to obtain
multipolar dynamos is multiplied by roughly $300$ when decreasing $E$ from
$10^{-4}$ to $10^{-5}$.  For each Ekman number, the width of the dipolar window
strongly depends on $\powerTrel$ since the actual input power needed to reach
the transition is an order of magnitude lower for pure chemical convection than
for pure thermal convection.  Simulations with $\powerTrel = 40\,\%$ are found
to behave similarly as purely thermal convection.  We further observe that
Earth-like dynamos can be obtained for any partitioning of power injection with
the best agreement obtained close to the dipole-multipole transition.  This is
a consequence of the way we have sampled the parameter space, mainly adopting
one single $Pm$ value for each Ekman number. Magnetic Reynolds numbers $Rm\sim
\mathcal{O}(1000)$ conducive to yield Earth-like fields are then attained at
strong convective forcings.  Adopting larger $Pm$ values at more moderate
chemical and thermal Rayleigh numbers could hence produce Earth-like fields
further away from the dipole-multipole transition
\cite[see][]{Christensen2010}.  Additional diagnostics are hence required to
better understand why the dipole-multipole transition depends so strongly on
the nature of the convective forcing.

%% file: dipole_multipole.tex
\subsection{Breakdown of the dipole}
\label{breakdown}
The physical reasons which cause the breakdown of the dipole in numerical
models remain poorly known.  Several previous studies suggest that the dipole
may collapse when inertia reaches a sizeable contribution in the force balance
\citep[see][]{Sreenivasan2006,Christensen2006}. 

\begin{figure}
	\centering
	\includegraphics[width=\linewidth]{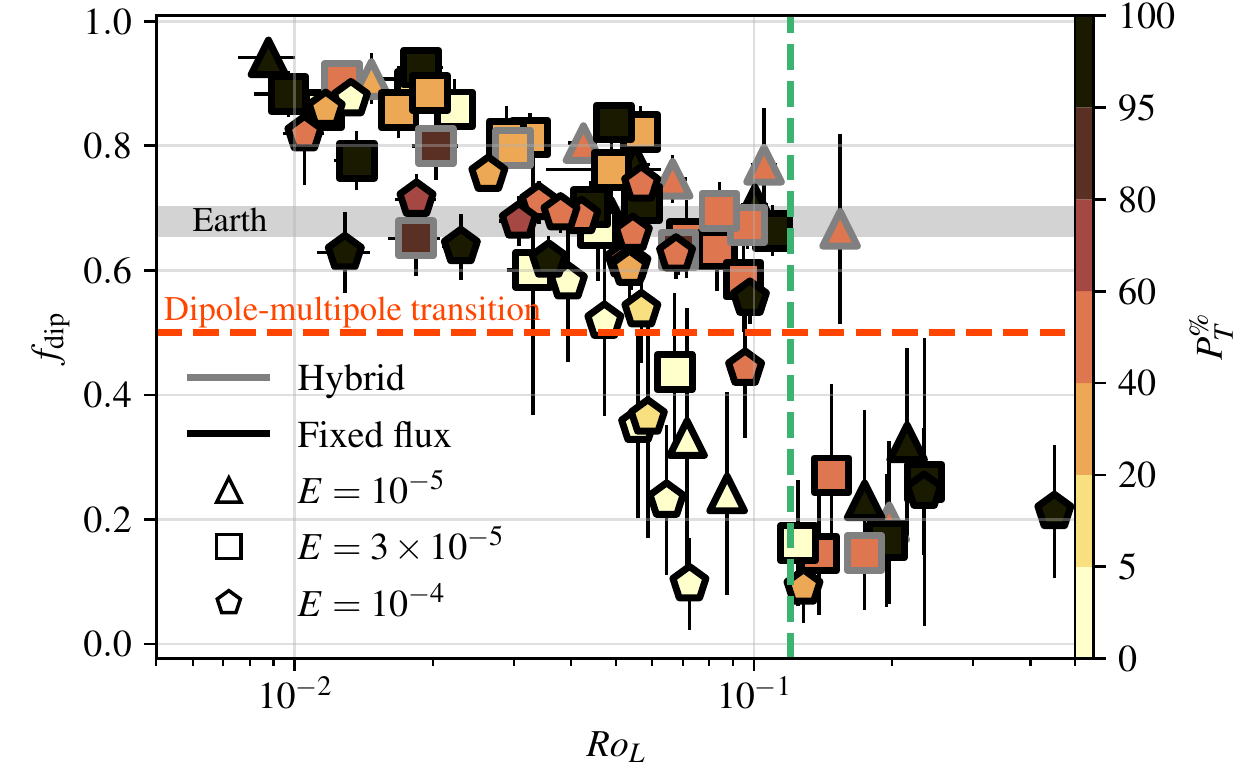}
	\caption{The dipolar fraction $\fdip$ as a function of the local
		Rossby number $\rol$. Geophysical range of $\fdip$ based on 
		the COV-OBS.x1 model by \cite{Gillet2015}. Black markers 
		correspond to simulations with $\powerT^\% = 100\,\%$, 
		while white ones correspond to simulations with $\powerT^\% = 0\,\%$. 
		The horizontal dashed line marks the limit between dipolar 
		and multipolar dynamos adopted in this study (see Sec.~\ref{sim_dia} 
		for details). The vertical dashed line marks the expected 
	 	limit between dipolar and multipolar dynamos according to 
		\cite{Christensen2006}. Vertical and horizontal black segments 
		attached to each symbol represent one standard deviation about 
		the time-averaged values.}
	\label{christensen}
\end{figure}

\cite{Christensen2006} introduced the local Rossby number $Ro_L$ as a proxy of
the ratio between inertia and Coriolis force and found no dipole-dominated
dynamos for $Ro_L > 0.12$ \citep[see][]{Christensen2010a}.
Figure~\ref{christensen} shows $\fdip$ as a function of $Ro_L$ for the $79$
simulations computed in this study.  The vertical line marks the threshold
value of $\rol = 0.12$ put forward by \cite{Christensen2006}, while the
horizontal line corresponds to $\fdip=0.5$, the boundary between dipolar and
multipolar dynamos adopted in this study.  To single out the effect of
partitioning the input power between chemical and thermal forcings, the symbols
have been color coded according to $\powerTrel$.  Each subset of models with
comparable $\powerTrel$ exhibits the same decrease of $\fdip$ with $\rol$ as
reported by \cite{Christensen2006}.  However, the dipole-multipole transition
occurs at lower $Ro_L$ when $\powerTrel$ decreases.  For $\powerTrel \geq
40\,\%$, the transition happens close to $\rol=0.12$ while it happens around
$\rol = 0.05$ for $\powerTrel = 0\,\%$.  In addition, the dynamo models with
$\powerTrel \geq 40\,\%$ are clearly separated in two groups of simulations
with either $\fdip \geq 0.5$ or $\fdip < 0.3$, while the dipole-multipole
transition is much more gradual for pure chemical forcing. $\rol$ hence fails
to capture the transition between dipolar and multipolar dynamos, independently
of the transport properties of the convecting fluid.

\begin{figure}
	\centering
	\includegraphics[width=\linewidth]{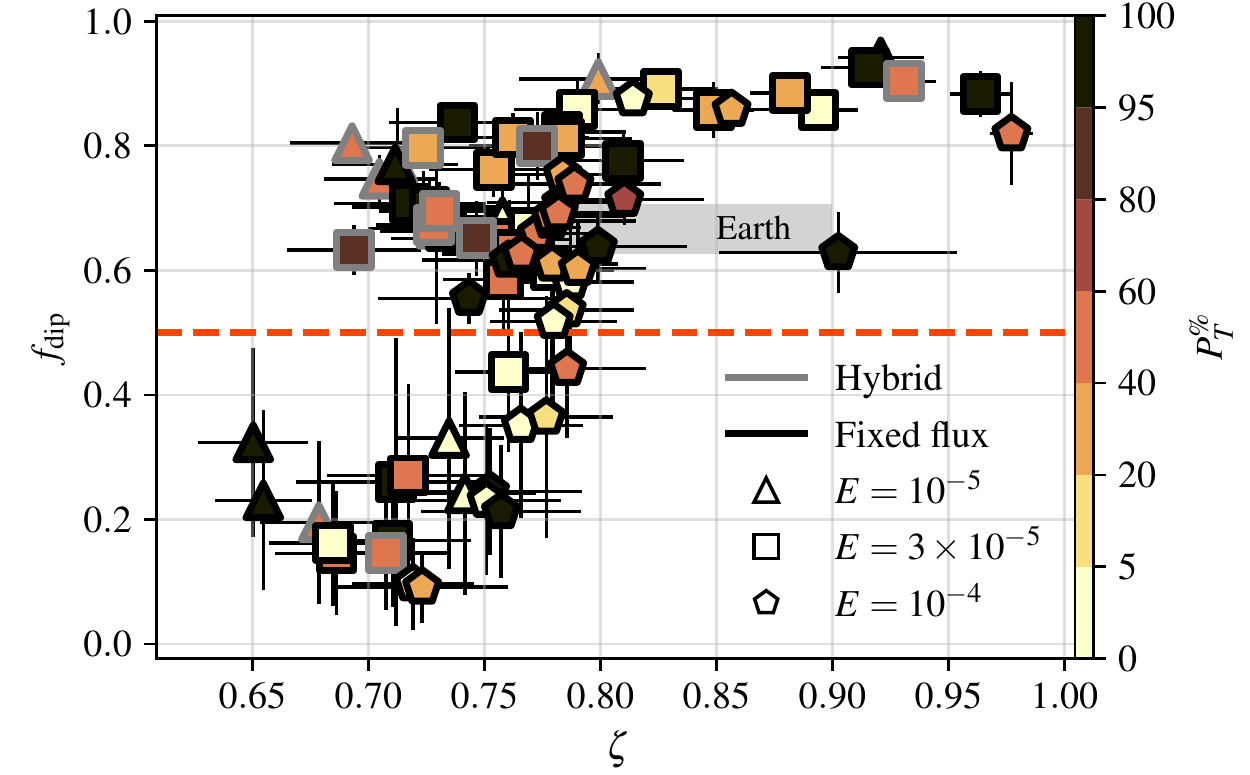}
	\caption{The dipolar fraction $\fdip$ as a function of the 
		relative portion of equatorially-symmetric kinetic energy $\ekrel$ 
		for the $79$ simulations computed in this study.  
		Geophysical range of $\fdip$  based on the COV-OBS.x1 model 
		by \cite{Gillet2015}. Geophysical estimates of $\ekrel$ are 
		based on the study of \cite{Aubert2014}. Black markers 
		correspond to simulations with $\powerT^\% = 100\,\%$, while 
		white ones correspond to simulations with $\powerT^\% = 0\,\%$. 
		The horizontal dashed line marks the limit between 
		dipolar and multipolar dynamos adopted in this
		study (see Sec.~\ref{sim_dia} for details). 
		Vertical and horizontal black segments attached to each symbol 
		represent one standard deviation about the time-averaged values.}
	\label{garcia}
\end{figure}

Following \cite{Christensen2006}, \cite{Garcia2017} also envision that the
increasing role of inertia would be responsible for the dipole breakdown. They
however define another parameter to characterize it.  They suggest that the
transition is related to a change in the equatorial symmetry properties of the
convective flow.  To quantify it, they introduce the ratio $\ekrel$ previously
defined in Eq.~(\ref{ekrel}).  Figure~\ref{garcia} shows  $\fdip$ as a function
of~$\ekrel$.  Geophysical estimates for $\ekrel$ are based on the study of
\cite{Aubert2014}.  The decrease of the relative equatorial symmetry $\ekrel$
goes along with a gradual weakening of the axisymmetric dipolar field.  Below
$\ekrel = 0.7$, no dipolar dynamos are obtained while conversely the models
with $\ekrel \geq 0.85$ are all dipolar.  However, this parameter has little
predictive power to separate the dipolar solutions from the multipolar ones
over the intermediate range $0.7 \leq \ekrel \leq 0.85$.

\begin{figure*}
	\centering
	\includegraphics[width=\linewidth]{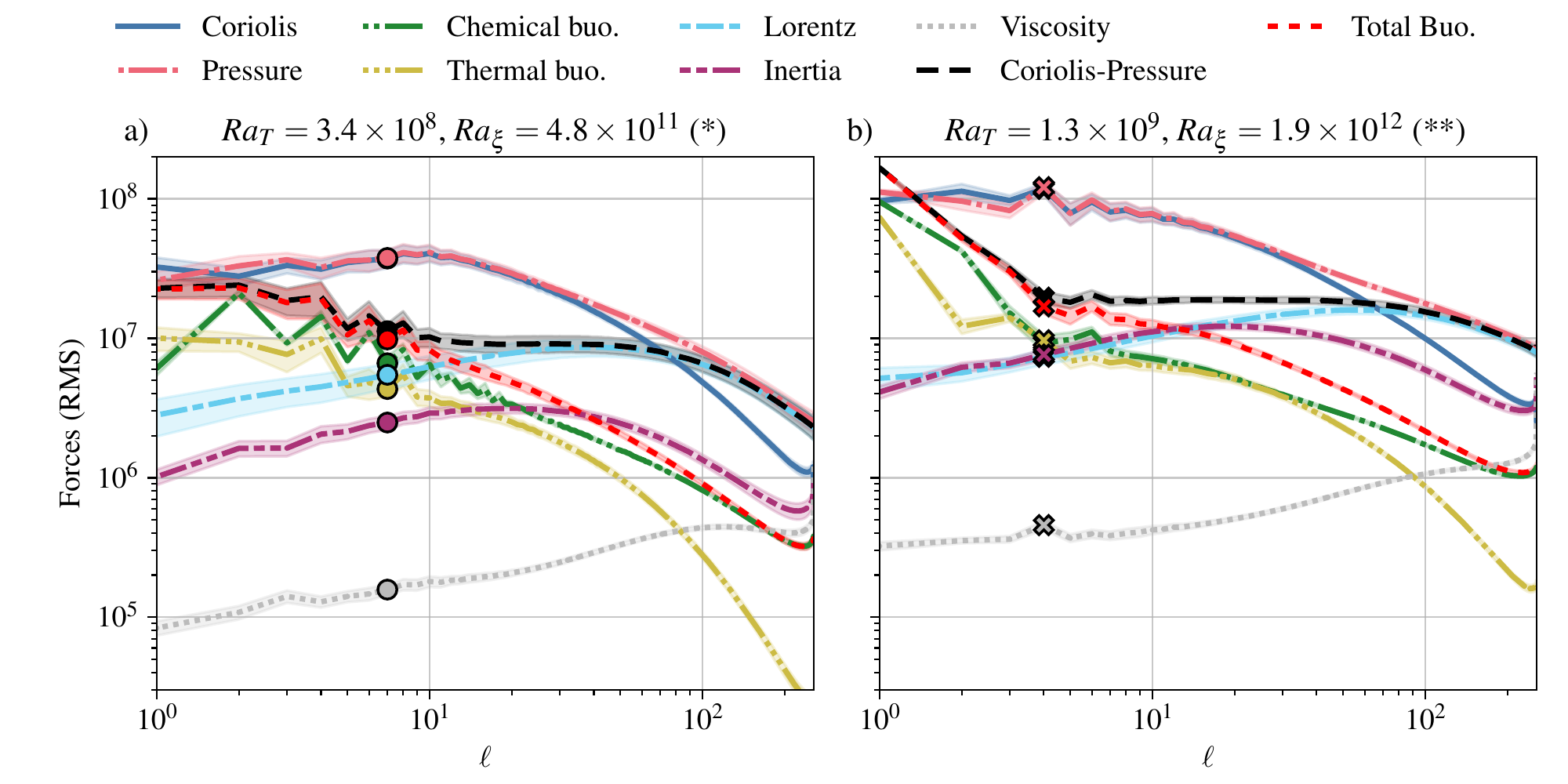}
	\caption{Force balance spectra as a function of the spherical harmonic 
		degree $\ell$ for a dipolar (a) and a multipolar (b) simulations 
		with $E = 10^{-5}$ and $\powerT^\% \approx 46\,\%$. 
		Thick lines correspond to the time average of each force, 
		while the shaded regions represent one standard deviation about mean.
		The abscissa of the markers corresponds to the dominant 
		lengthscale $\lpic$ for each simulation. A circle corresponds 
		to a dipolar simulation while a cross corresponds to a 
		multipolar one. Both simulations are referenced as simulations 
		(*) and (**) in Tab.~\ref{simu_tab}.}
	\label{spectre}
\end{figure*}

Another way to examine the dipole-multipole transition resorts to looking at
the force balance governing the outer core flow dynamics
\citep[see][]{Soderlund2012,Soderlund2015}.  To do so, we analyse force balance
spectra following \cite{Aubert2017} and \cite{Schwaiger2019}.  Each force
entering the Navier-Stokes equation~(\ref{NS_dim}) is hence decomposed into
spherical harmonic contributions.  The spatial root mean square
$F_\mathrm{RMS}$ of a vector $\mathbf{F}$ reads

\begin{equation}
	F_\mathrm{RMS}(t) = \sqrt{\left\langle \mathbf{F}^2(\mathbf{r}, t)
	\right\rangle_{V_{o \setminus \delta}}},
	\label{RMS}
\end{equation}
where $\delta$ represents the thickness of the viscous boundary layer and $V_{o
\setminus \delta}$ the outer core volume that excludes those boundary layers.
By using the decomposition in spherical harmonics, the above expression can be
rearranged as
\begin{equation*}
		F^2_\mathrm{RMS}(t) = \dfrac{1}{V_{o \setminus \delta}} 
		\sum_{\ell=0}^{\ell_\mathrm{max}}\displaystyle \int_{r_i + \delta}^{r_o - 
\delta}\sum_{m = -\ell}^{\ell} |F_{\ell m}(r, t)|^2 r^2 \mathrm{d}r,
\end{equation*}
We define the time-averaged spectrum $F^{\ell}$ as a function of the harmonic 
degree $\ell$ for the force $\mathbf{F}$ by the relation

\begin{equation}
	{F^\ell} = \overline{\sqrt{\dfrac{1}{V_{o\setminus \delta}} \displaystyle 
\int_{r_i + \delta}^{r_o - \delta}\sum_{m = -\ell}^{\ell} |F_{\ell 
m}(r, t)|^2 r^2 \mathrm{d}r}}\,.
\end{equation}

\begin{figure}
	\centering
	\includegraphics[width=\linewidth]{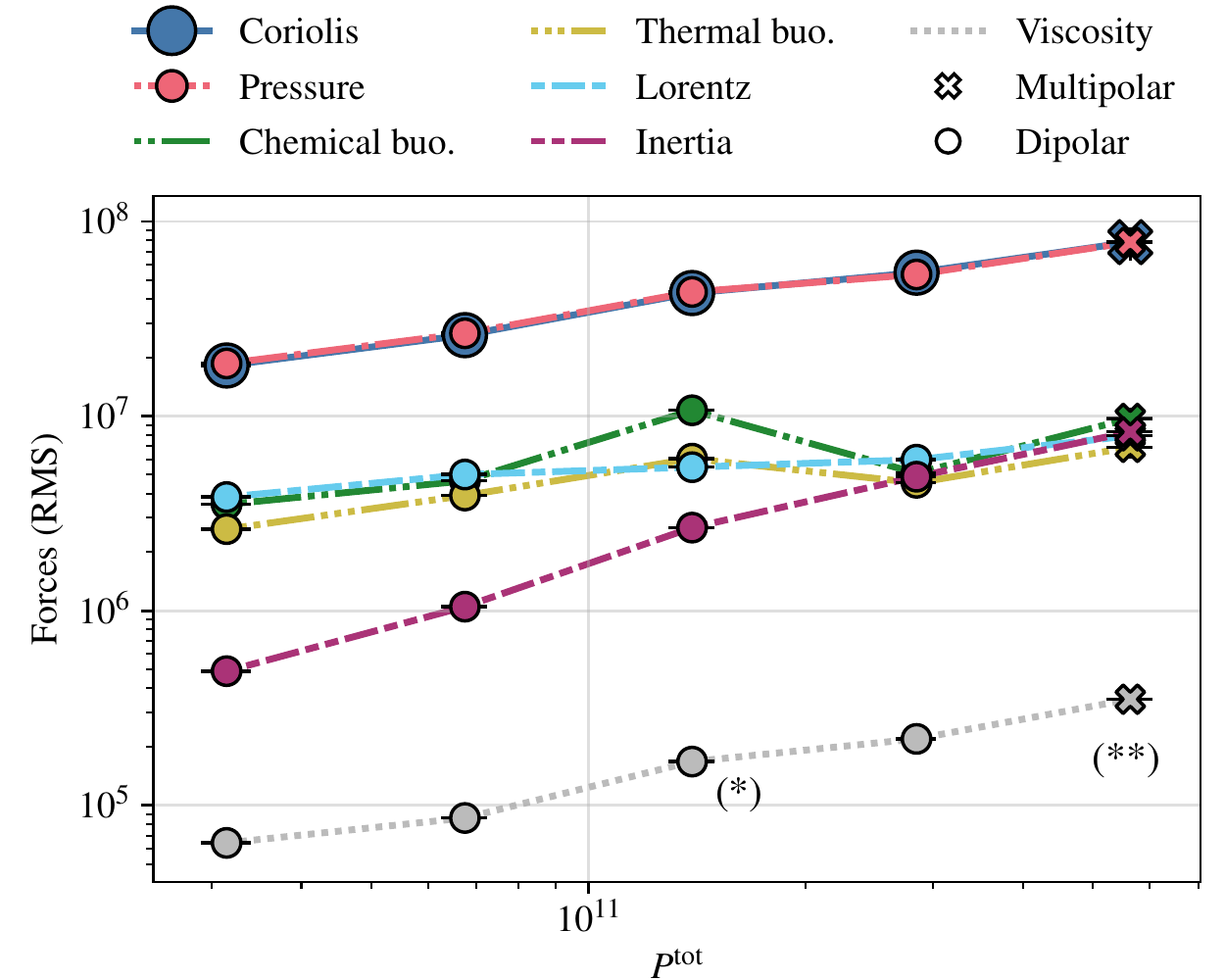}
	\caption{Time-averaged force balance spectra at the dominant
		lengthscale $\lpic$ as a function the 
		total buoyancy power $\powertot$ for numerical models with 
		$E = 10^{-5}$ and $30\,\%<\powerT^\%<60\,\%$. 
		The dynamo models (*) and (**) correspond to simulations 
		referenced in Tab.~\ref{simu_tab}. The horizontal and vertical 
		segments attached to each symbol correspond to one standard 
		deviation about the time-averaged values.}
	\label{soderlund}
\end{figure}
Figure~\ref{spectre} shows the time-averaged force balance spectra for one
dipolar and one multipolar dynamo with $E=10^{-5}$ and $\powerT\%\simeq 46\%$.
For both panels, the spherical harmonic at which the poloidal kinetic energy
peaks $\lpic$ is indicated by filled markers.

The left panel corresponds to the force balance of the reference case (*) which
is in excellent agreement with the geomagnetic field in terms of its low
$\chi^2$ value (recall Sec.\ref{ref_case}).  Its spectra feature a dominant
quasi-geostrophic balance between Coriolis and pressure forces up to $\ell
\approx 60$ accompanied by a magnetostrophic balance at smaller scales.  The
difference between pressure and Coriolis forces, forming the so-called
ageostrophic Coriolis force (long dashed line), is then balanced by the two
buoyancy sources (short dashed line) at large scales and by Lorentz force
(irregular dashed line) at small scales.  This forms the
quasi-geostrophic Magneto-Archimedean-Coriolis balance (QG-MAC) devised by
\cite{Davidson2013} and expected to control the outer core fluid dynamics
\citep[see][]{Roberts2013}. This hierarchy of forces is similar to the one
observed in geodynamo models that use a codensity approach
\citep[e.g.][]{Aubert2017,Schwaiger2019}.  The breakdown of buoyancy sources
reveals a dominant contribution of chemical forcings which grows at small
scales.  The QG-MAC balance is perturbed by a sizeable inertia, which reaches
almost a third of the amplitude of Lorentz force below $\ell \approx 20$, while
viscous effects are deferred to more than one order of magnitude below. Because
of the strong convective forcing, the force separation is hence not as
pronounced as in the exemplary dipolar cases by \citep[][their
Fig.~2]{Schwaiger2019}.

\begin{figure*}
	\centering
	\includegraphics[width=\linewidth]{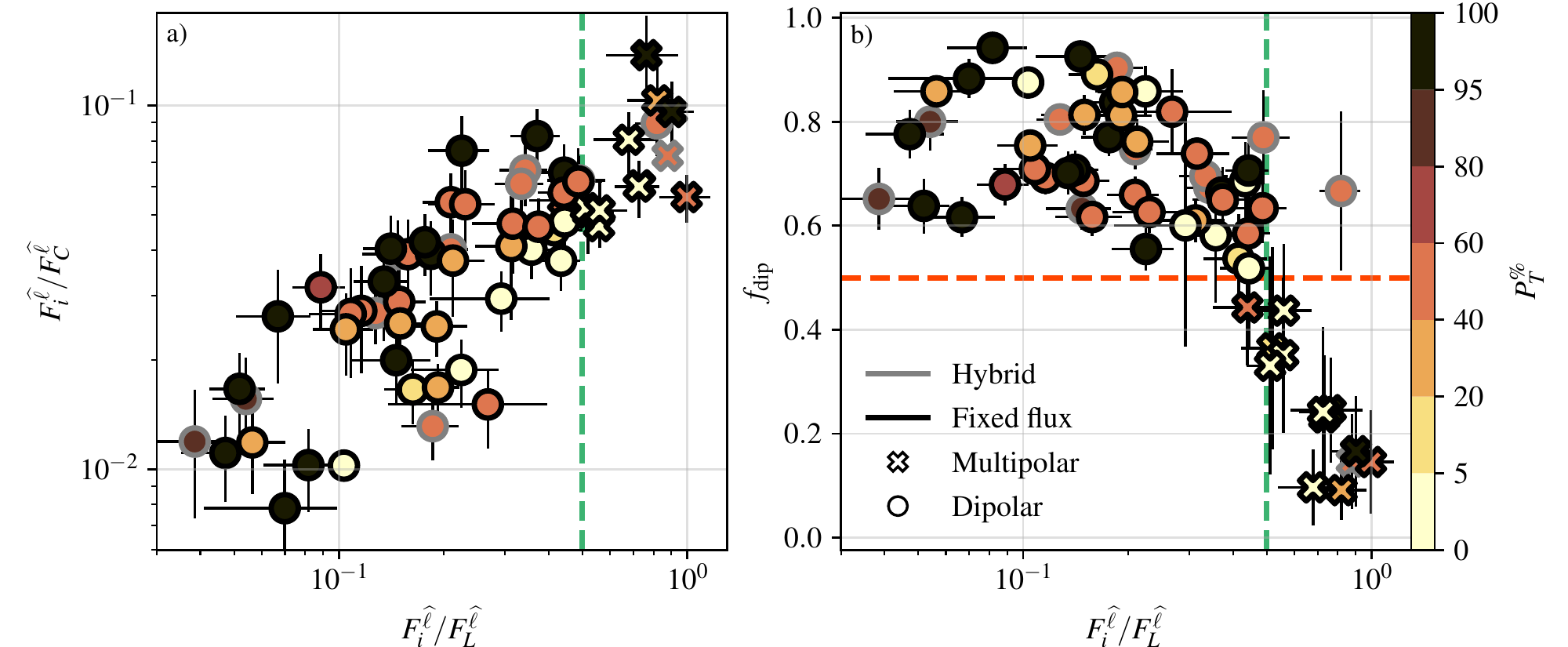}
	\caption{Ratio of inertia to Coriolis force at the dominant lengthscale 
	$F_i^{\lpic} / F_C^{\lpic}$ (left) and the dipolar fraction $\fdip$ (right) 
	as a function of 
	the ratio of inertia to the Lorentz force at the dominant lengthscale 
	$F_i^{\lpic} / F_L^{\lpic}$. Black markers correspond 
	to simulations with $\powerT^\% = 100\,\%$, while white 
	markers correspond to simulations with $\powerT^\% = 0\,\%$. 
	Horizontal dashed line marks the limit between 
	dipolar and multipolar dynamo according to Sec.~\ref{sim_dia}. 
	Vertical dashed line marks the limit between 
	dipolar and multipolar dynamo in term of $F_i^{\lpic} / F_L^{\lpic}$ ratio. 
	Vertical and horizontal black segments attached to the symbol 
        correspond to one standard deviation about the time-averaged values.}
	\label{menu}
	
	\includegraphics[width=0.8\linewidth]{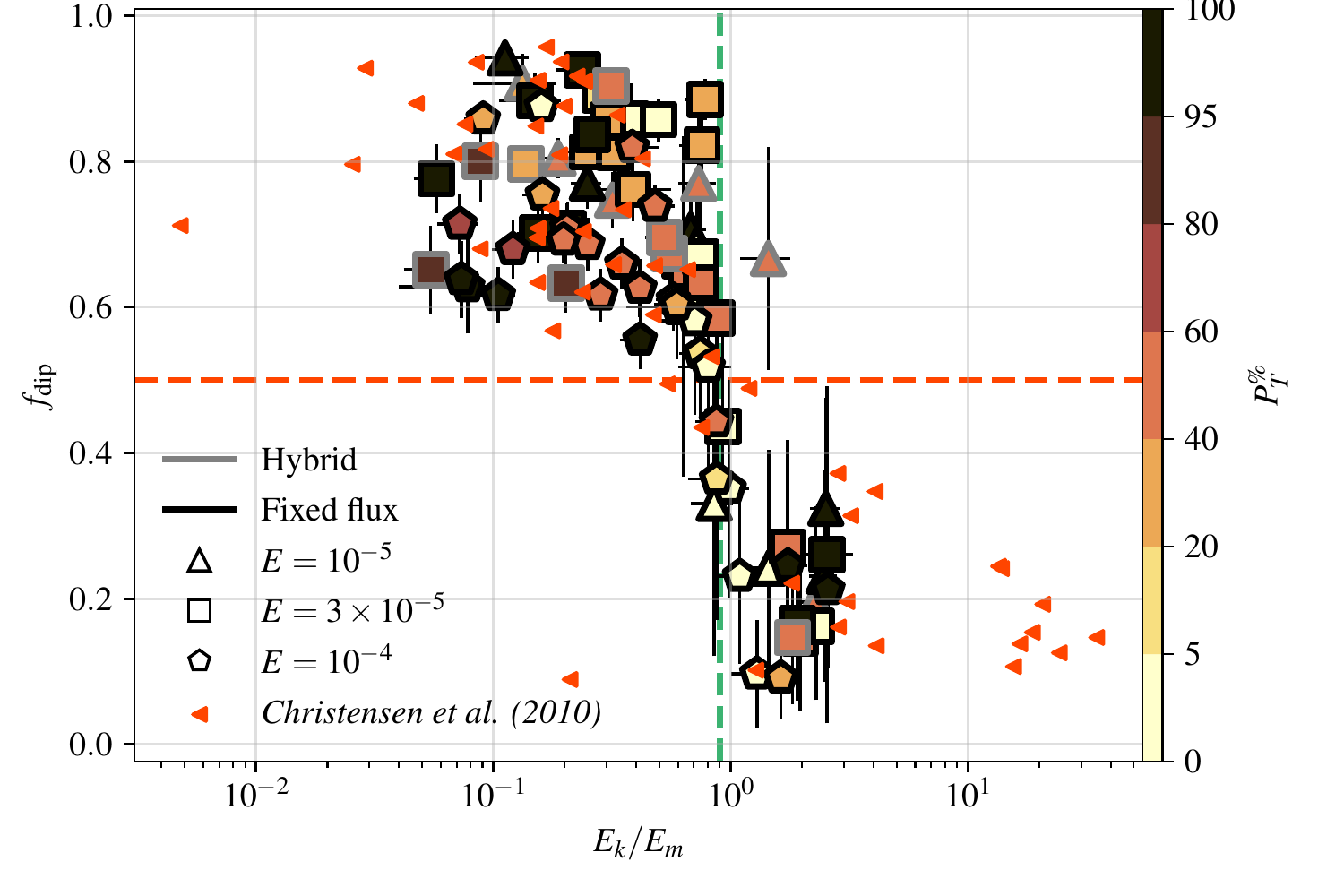}
	\caption{Dipolarity parameter $\fdip$ as a function of the kinetic to magnetic 
	        energy ratio $E_k/E_m$. 
		The triangles correspond to 
		those simulations by \cite{Christensen2010a}
		with $Pr \neq 1$.
		The horizontal dashed line marks the 
		$\fdip=0.5$ limit between 
		dipolar and multipolar dynamos. 
		The vertical dashed line corresponds to $E_k/E_m=0.9$.
		Vertical and horizontal black segments attached to the 
		symbols correspond to one standard deviation about the time-averaged 
values for $\fdip$ and $E_k/E_m$, respectively.}
	\label{ek_em}
\end{figure*}

By increasing the input power by a factor $4$ while keeping $\powerTrel$
constant, we obtain a numerical model (**) (see Tab~\ref{simu_tab} and
Fig.~\ref{ra_raxi}c) where dynamo action yields a multipolar field.  As
compared to the dipole-dominated solution, the amplitude of each contribution
is hence shifted to higher values.  Most noticeable changes concern the
prominent contribution of the chemical buoyancy for the degree $\ell=1$ and the
ratio of inertia to Lorentz force.  The former comes from a pronounced
equatorial asymmetry of the chemical fluctuations. The development of strong
equatorially-asymmetric convective motions has been observed by
\cite{Landeau2011} and \cite{Dietrich2013} with a codensity approach and flux
boundary conditions.  Below $\ell \approx 20$, inertia reaches a comparable
amplitude to Lorentz force, while the smaller scales are still controlled by
magnetic effects.  This differs from the multipolar dynamo model described by
\cite{Schwaiger2019}, where inertia was significantly larger than Lorentz force
at all scales forming the so-called quasi-geostrophic
Coriolis-Inertia-Archimedian balance \citep[e.g.][]{Gillet2006}.  Here the
situation differs likely because of the larger $Pm$, which enables a stronger
magnetic field \citep[see][]{Menu2020}.  At the dominant lengthscale $\lpic$,
the ratio $F_i^{\ell} / F_L^{\ell}$ is around $1$ for the multipolar model,
while it is less than $0.5$ for the dipolar one. To examine whether the
dipole-multipole transition is controlled by the ratio of inertia over Lorentz
forces, we hence focus on the force balance at the dominant lengthscale
$\lpic$, in contrast to previous studies, which analysed ratio of integrated
forces \citep{Soderlund2012,Soderlund2015,Yadav2016}. 

Figure~\ref{soderlund} shows the time-averaged force balance at $\lpic$ for the
simulations with $E = 10^{-5}$ and $ 30\,\% \leq \powerT^\% \leq 60\,\%$.  The
dynamics at $\lpic$ is primarily controlled by the geostrophic balance between
the Coriolis force and the pressure gradient.  The other contributions grow
differently with $\powertot$: viscosity and  inertia increase continuously
while Lorentz force at $\lpic$ hardly increases beyond $\powertot \approx 3
\times 10^{10}$.  The dipole-multipole transition occurs when inertia reaches a
comparable amplitude to Lorentz force at $\lpic$ (crosses).

The relevance of this force ratio for sustaining the dipolar field has already
been put forward by \cite{Menu2020}, using models with a purely thermal forcing
and $Pr=1$.  By considering turbulent simulations with large $Pm$, they show
that strong Lorentz forces at large scale prevent the collapse of the dipole by
inertia.  As a result, they report dipole-dominated simulations with $Ro_L$
which exceeds the limit of $0.12$ proposed by \cite{Christensen2006}.  Here, we
quantify the contribution of inertia at the dominant convective  lengthscale by
dividing the amplitude of inertia $F_i^{\lpic}$ by the amplitude of Coriolis
$F_C^{\lpic}$ and Lorentz $F_L^{\lpic}$ forces.

Figure~\ref{menu}(a) shows our simulations in the parameter space defined by
the amplitude ratios ($F_i^{\lpic} / F_L^{\lpic}$, $F_i^{\lpic} /
F_C^{\lpic}$).  Each simulation is characterized by the proportion of thermal
convection $\powerT^\%$ and the nature of its thermal boundary conditions.
Increasing the input power of the dynamo leads to a growth of inertia, such
that the strongly-driven cases all lie in the upper right quadrant of
Fig.~\ref{menu}(a).  The transition from dipolar to multipolar dynamos occurs
sharply when $F_i^{\lpic} / F_L^{\lpic}$ exceeds $0.5$ over a broad range of
$F_i^{\lpic} / F_C^{\lpic}$ ranging from $0.03$ to $0.08$.  This indicates that
the transition is much more sensitive to the ratio of inertia over Lorentz
force than to the ratio of inertia over Coriolis force.  The transition for
purely chemical simulations (white symbols) is reached at lower values of
$F_i^{\lpic} / F_C^{\lpic}$ and is more continuous than for the thermal ones
(black symbols).  This confirms the trend already observed in
Fig.~\ref{christensen}, where $\fdip$ shows a much more gradual decreases with
$\rol$ when $\powerT^\%=0\%$.

Figure~\ref{menu}(b) shows $\fdip$ as a function of $F_i^{\lpic} /
F_L^{\lpic}$.  In contrast with the previous criteria, the ratio $F_i^{\lpic} /
F_L^{\lpic}$ successfully captures the transition between dipole-dominated and
multipolar dynamos which happens when $F_i^{\lpic} / F_L^{\lpic} \simeq 0.5$,
independently of the buoyancy power fraction.  The simulation, which singles
out in the upper right quadrant of Fig.~\ref{menu}(b), is an exception to this
criterion.  This numerical model corresponds to $\powerT^\% \approx 47\,\%$
with $\rat = 6.8 \times 10^{10}$, $\rac = 9.6 \times 10^{11}$ with hybrid
boundary conditions. Although it features $F_i^{\lpic}/F_L^{\lpic} > 0.5$, its
magnetic field is on time-average dominated by an axisymmetric dipole ($\fdip =
0.69$). This dynamo however strongly varies with time with several drops of the
dipolar component below $\fdip=0.5$ (see Fig.~\ref{fdip_68} in the appendix).
Although the numerical model has been integrated for more than two magnetic
diffusion times, the stability of the dipole cannot be granted for certain.

%% file: discussion.tex
\section{Summary and discussion}
\label{discussion}
Convection in the liquid outer core  of the Earth is thought to be driven by
density perturbations from both thermal and chemical origins.  In the vast
majority of geodynamo models, the difference between the two buoyancy sources
is simply ignored.  In planetary interiors with huge Reynolds numbers,
diffusion processes associated with molecular diffusivities could indeed
possibly be superseded by turbulent eddy diffusion \citep[see][]
{Braginsky1995}. This hypothesis forms the backbone of the so-called
``codensity'' approach which assumes that both thermal and compositional
diffusivities are effectively equal.  This approach suppresses some  dynamical
regimes intrinsic to double-diffusive convection \citep{Radko2013}.

The main goal of this study is to examine the impact of double-diffusive
convection on the magnetic field generation when both thermal and compositional
gradients are destabilizing \citep[the so-called top-heavy regime,
see][]{Takahashi2014}. To do so we have computed $79$ global dynamo models,
varying the fraction between thermal and compositional buoyancy sources
$\powerT^\%$, the Ekman number $E$ and the vigor of the convective forcing
using a Prandtl number $Pr=0.3$ and a Schmidt number $Sc=3$. We have explored
the influence of the thermal boundary conditions by considering two sets of
boundary conditions for temperature and composition.

Using a generalised eigenvalue solver, we have first investigated the onset of
thermo-solutal convection.  In agreement with previous studies
\citep{Busse2002,Net2012,Trumper2012,Silva2019}, we have shown that the
incorporation of a destabilizing compositional gradient actually facilitates
the onset of convection as compared to the single diffusive configurations by
reducing the critical thermal Rayleigh number. The critical onset mode in the
top-heavy regime of rotating double-diffusive convection is otherwise similar
to classical thermal Rossby waves obtained in purely thermal convection
\citep{Busse1970}.

To quantify the Earth-likeness of the magnetic fields produced by the
non-linear dynamo models, we have used the rating parameters introduced by
\cite{Christensen2010}.  Using geodynamo models with a codensity approach,
\cite{Christensen2010} suggested that the Earth-like dynamo models are located
in a wedge-like shape in the 2-D parameter space constructed from the ratio of
three typical timescales, namely the rotation rate, the turnover time and the
magnetic diffusion time.  Here, we have shown that the physical parameters at
which the best morphological agreement with the geomagnetic field is attained
strongly depend on the ratio of thermal and compositional input power. In
particular, we obtain $6$ purely-compositional multipolar dynamo models that
lie within the wedge region of Earth-like dynamos (recall Figure~\ref{wedge}a).
This questions the relevance of the regime boundaries proposed by
\cite{Christensen2010}.

We have then used our set of double-diffusive dynamos to examine the transition
between dipolar and multipolar dynamos. We have assessed the robustness of
several criteria controlling this transition that had been proposed in previous
studies.  \cite{Sreenivasan2006} suggested that the dipole breakdown results
from an increasing role played by inertia at strong convective forcings.
\cite{Christensen2006} then introduced the local Rossby number $\rol$ as a
proxy of the ratio of inertial to Coriolis forces. They suggested that
$\rol\approx 0.12$ marks the boundary between dipole-dominated and multipolar
dynamos over a broad range of control parameters.  Our numerical dynamo models
with $\powerTrel \geq 40\%$ follow a similar behaviour, while the transition
between dipolar and multipolar dynamos occurs at lower $\rol$ ($\approx 0.05$)
when chemical forcing prevails. A breakdown of the dipole for dynamo models
with $\rol < 0.1$ was already reported by \cite{Garcia2017} using $Pr >1$ under
the codensity hypothesis (their Fig.~1). 

Using non-magnetic numerical models, \cite{Garcia2017} further argued that the
breakdown of the dipolar field is correlated with a change in the equatorial
symmetry properties of the convective flow.  They introduced the relative
proportion of kinetic energy contained in the equatorially-symmetric convective
flow, $\ekrel$, and suggested that multipolar dynamos would be associated with
a lower value of this quantity.  However, our numerical dataset shows that
multipolar and dipolar dynamos coexist over a broad range of $\ekrel$
($0.70-0.85$, recall Fig.~\ref{garcia}), indicating that this ratio has little
predictive power in separating dipolar from multipolar simulations. 

While neither $Ro_L$ nor $\ekrel$ provide a satisfactory measure to
characterise the dipole-multipole transition, the analysis of the force balance
governing the dynamo models has been found to be a more promising avenue to
decipher the physical processes at stake \citep{Soderlund2012,Soderlund2015}.
By considering a spectral decomposition of the different forces
\citep[e.g.][]{Aubert2017,Schwaiger2019}, we have shown that the transition
between dipolar and multipolar dynamos goes along with an increase of inertia
at large scales. The analysis of the force ratio at the dominant scale of
convection has revealed that the dipole-multipole transition is much more
sensitive to the ratio of inertia to Lorentz force than to the ratio of inertia
to Coriolis force. The transition from dipolar to multipolar dynamos robustly
happens when the ratio of inertial to magnetic forces at the dominant
lengthscale of convection exceeds $0.5$, independently of $\powerTrel$ and the
Ekman number. This confirms the results by \cite{Menu2020} who argued that a
strong Lorentz force prevents the demise of the axial dipole, delaying its
breakdown beyond $\rol\approx 0.12$ (their Figs.~3 and 4).

Providing a geophysical estimate of the ratio of inertial to magnetic forces at
the dominant scale of convection in the Earth's core is not an easy task.
Recent work by \cite{Schwaiger2020}  suggests that the dominant scale of
convection should be that at which the Lorentz force and the buoyancy force,
both second-order actors in the force balance, equilibrate.  Extrapolation of
this finding to Earth's core yields a scale of approximately $200$~km, that
corresponds to spherical harmonic degree~$40$. This is far beyond what can be
constrained through the analysis of the geomagnetic secular variation.
Estimating the strength of both Lorentz and inertial forces at that scale is
hence out of reach. 

We can however try to approximate the ratio of these two forces by a simpler
proxy, namely the ratio of the total  kinetic energy $E_k$ to total magnetic
energy $E_m$. To examine the validity of this approximation, Fig.~\ref{ek_em}
shows $\fdip$ as a function of the ratio $E_k/E_m$ for our numerical
simulations complemented with the codensity simulations of
\cite{Christensen2010a} that have $Pr \neq 1$.  We observe that the dipolar
fraction $\fdip$ exhibits a variation similar to that shown in
Fig.~\ref{menu}(b) for the actual force ratio. The transition between dipolar
and multipolar dynamos is hence adequately captured by the ratio $E_k/E_m$
\citep{Kutzner2002}. All but one of the numerical dynamos of our dataset become
multipolar for  $E_k/E_m > 0.9$, independently of $E$, $\powerTrel$, and the
type of thermal boundary conditions prescribed.  Using the physical properties
from Tab.~\ref{phy_param} leads to the following estimate for the Earth's core
\begin{equation*}
	\frac{E_k}{E_m} = \mu_0\rho_o\frac{u_\mathrm{rms}^2}{B_\mathrm{rms}^2} \approx 10^{-4} - 
10^{-3} \ll \mathcal{O}(1).
\end{equation*}
\cite{Christensen2010a} and \cite{Wicht2010} argued that convection in the
Earth's core should operate in the vicinity of the transition between dipolar
and multipolar dynamos in order to explain the reversals of the geomagnetic
field.  This statement, however, postulates that reversals  and the dipole
breakdown are governed by the same physical mechanism.  The smallness of
$E_k/E_m$ for Earth's core indicates that it should operate far from the
dipole-multipole transition, contrary to the numerical evidence accumulated up
to now. The paleomagnetic record indicates that during a reversal or an
excursion, the intensity of the field is remarkably low, which suggests that
the strength of the geomagnetic field could decrease by about an order of
magnitude. This state of affairs admittedly  brings the ratio $E_k/E_m$ closer
to unity, yet without reaching it. So it seems that the occurrence of
geomagnetic reversals is not directly related to an increase of the relative
amplitude of inertia. Other mechanisms proposed to explain geomagnetic
reversals rely on the interaction of a limited number of magnetic modes, whose
nonlinear evolution is further subject to random fluctuations
\citep[e.g.][]{schmitt2001magnetic,petrelis2009simple}.  In this useful
conceptual framework, the large-scale dynamics of Earth's magnetic field is
governed by the induction equation alone.  The origin of the fluctuations that
can potentially lead to a reversal of polarity is not explicited and it remains
to be found, but evidence from mean fields models by \cite{Frankie} suggests
that the likelihood of reversals increases with the magnetic Reynolds number.
In practice, these fluctuations could very well occur in the vicinity of the
convective lengh scale, and have either an hydrodynamic, or a magnetic, or an
hydromagnetic origin, depending on the process driving the instability.
Shedding light on the origin of these fluctuations constitutes an interesting
avenue for future numerical investigations of geomagnetic reversals.

%% file: acknowledgments.tex
We would like to thank Ulrich Christensen and Julien Aubert for sharing their
data.  Numerical computations were performed at S-CAPAD, IPGP and using HPC
resources from GENCI-CINES and TGCC-Irene KNL (Grant 2020-A0070410095).  Those
computations required roughly $3.3$ millions single core hours on Intel Haswell
CPUs. This amounts for $\unit{33}{\mega \watt \hour}$, or equivalently to a bit
more than $3$ tons of carbon dioxide emissions. We thank Gabriel Hautreux, from
GENCI-CINES, for providing us with a direct measure of the power consumption of
MagIC.  All the figures have been generated using \texttt{matplotlib}
\citep{Hunter2007} and paraview (\url{https://www.paraview.org}).  The colorbar
used in this study were designed by \cite{Thyng2016} and
\cite{Crameri2018,Crameri2019}.

%% file: appendix_time.tex
\section{Time scheme}
\label{time-scheme}
We provide in this appendix 
the matrices $\mathbf{a}^\mathcal{I}$ and
$\mathbf{a}^\mathcal{E}$ and the vectors $\mathbf{b}^\mathcal{I}$,
$\mathbf{b}^\mathcal{E}$, $\mathbf{c}^\mathcal{E}$ and $\mathbf{c}^\mathcal{I}$
of the two IMEX Runge-Kutta schemes that we resorted to for this study. 
These vectors and matrices can conveniently be represented using 
Butcher tables \citep[][]{butcher1964runge}. 
\renewcommand{\arraystretch}{1.2}

\vspace{2em}

\begin{minipage}[t]{0.35\linewidth}
	\textbf{BPR353:}  \citep{Boscarino2013}

	Implicit component
	\begin{equation*}
		\begin{array}[c]{c|c}
			\mathbf{c}^\mathcal{I} & \mathbf{A}^\mathcal{I}\\ 
			\hline 
					       & \mathbf{b}^\mathcal{I}
		\end{array}
		=
		\begin{array}[c]{c|c c c c c}
			0    & 0    &         &       &   & \\
			1    & 1/2    & 1/2       &       &   & \\
			2/3  & 5/18  & -1/9     & 1/2     &   & \\
			1    & 1/2  & 0       & 0   &1/2  & \\
			1    & 1/4  & 0       & 3/4   & -1/2 & 1/2\\\hline
			     & 1/4  & 0       & 3/4   & -1/2 & 1/2
		\end{array}
	\end{equation*}
	Explicit component
	\begin{equation*}
		\begin{array}[c]{c|c}
			\mathbf{c}^\mathcal{E} & \mathbf{A}^\mathcal{E}\\ 
			\hline 
					       & \mathbf{b}^\mathcal{E}
		\end{array}
		=
		\begin{array}[c]{c|c c c c c}
			0    & 0    &         &       &   & \\
			1    & 1    & 0       &       &   & \\
			2/3  & 4/9  & 2/9     & 0     &   & \\
			1    & 1/4  & 0       & 3/4   &0  & \\
			1    & 1/4  & 0       & 3/4   & 0 & 0\\\hline
			     & 1/4  & 0       & 3/4   & 0 & 0
		\end{array}
	\end{equation*}
\end{minipage}
\hfill ~
\begin{minipage}[t]{0.3\linewidth}
	\textbf{PC2:} \citep{Jameson1981}

	Implicit component
	\begin{equation*}
		\begin{array}[c]{c|c}
			\mathbf{c}^\mathcal{I} & \mathbf{A}^\mathcal{I}\\ 
			\hline 
					       & \mathbf{b}^\mathcal{I}
		\end{array}
		=
		\begin{array}[c]{c|c c c c }
			0    & 0  &      &      &\\
			1    & 1/2  & 1/2       &    &\\
			1    & 1/2  & 0       & 1/2\\
			1    & 1/2  & 0       & 0 & 1/2\\\hline
			     & 1/2  & 0       & 0   & 1/2
		\end{array}
	\end{equation*}
	Explicit component
	\begin{equation*}
		\begin{array}[c]{c|c}
			\mathbf{c}^\mathcal{E} & \mathbf{A}^\mathcal{E}\\ 
			\hline 
					       & \mathbf{b}^\mathcal{E}
		\end{array}
		=
		\begin{array}[c]{c|c c c c }
			0    & 0  &      &      &\\
			1    & 1  & 0       &    &\\
			1    & 1/2  & 1/2       & 0\\
			1    & 1/2  & 0       & 1/2 & 0\\\hline
			     & 1/2  & 0       & 1/2   & 0
		\end{array}
	\end{equation*}
\end{minipage}

%% file: table.tex
\section{Numerical simulations}
{\small
\renewcommand{\arraystretch}{1.1}
\begin{longtable}{p{0.1cm} p{0.5cm} p{0.5cm} p{1.3cm} p{0.4cm} p{0.5cm} p{0.8cm} 
	          *{3}{p{0.4cm}} p{0.7cm} *{8}{p{0.4cm}} p{0.1cm}}
\kill
\caption{Control parameters and simulation diagnostics for the $79$ numerical 
simulations computed for this study. 
	 Simulations computed using the finite difference method in radius
	 are marked with a superscript $\mathrm{f}$ (the others were computed using the Chebyshev collocation method
	 in radius). 
Simulations with hybrid boundary conditions are marked by an H in the 
first column.
Simulations are sorted by growing Ekman number and then by growing magnetic Reynolds number. 
The averaging and running times 
$t_\mathrm{avg}$ and $t_\mathrm{run}$ are expressed in units of magnetic diffusion time $\tau_\eta$.}\\ 
\label{simu_tab}
&$\rat$ & $\rac$             & $(n_r, \ell_\mathrm{max})$ & $Pm$ & $\alpha$ & $t_\mathrm{scheme}$
       & $t_\mathrm{avg}$   & $t_\mathrm{run}$ & $Rm$ & $\rol$ & $\Lambda$ & $f_\mathrm{ohm}$
       & $\chi^2$ & $\zeta$ & $f_\mathrm{dip}$ & $\powerTrel$ & $Nu$ & $Sh$ &\\
       &$(\times 10^{8})$ & $(\times 10^{9})$ &&&&&&&& $(\times 10^{-2})$\\       \hline
\endfirsthead
\caption{Control parameters and simulation diagnostics for the 79 numerical simulations computed for this study (continued).}\\
&$\rat$ & $\rac$& $ (n_r, \ell_\mathrm{max})$ & $Pm$ & $\alpha$ & $t_\mathrm{scheme}$
       & $t_\mathrm{avg}$ & $t_\mathrm{run}$ & $Rm$ & $\rol$& $ \Lambda$ & $f_\mathrm{ohm}$
       &$\chi^2$ & $\zeta$ & $f_\mathrm{dip}$ & $\powerTrel$ & $Nu$ & $Sh$ & \\
       &$(\times 10^{8})$ & $(\times 10^{9})$&&&&&&&& $(\times 10^{-2})$\\       \hline
\endhead
\multicolumn{20}{c}{$E = 1 \times 10^{-5},\quad Pr = 0.3,\quad Sc = 3$} \\\hline
H& $0.1$  &  $10$  &  $(97, 133)$ & $1.0$ & $0.935$ & BPR353 & $1.80$ & $1.80$ & $176$ & $1.5$ & $2.4$ & $0.66$ & $20.1$ & $0.80$ & $0.91$ & $32$ & $1.8$ & $10.6$\\
& $3.1$ & $0$ & $(97, 170)$ & $1.0$ & $0.935$ & BPR353 & $2.79$ & $3.09$ & $196$ & $0.9$ & $3.4$ & $0.79$ & $12.6$ & $0.92$ & $0.94$ & $100$ & $2.0$ & $1.0$\\
& $0$ & $100$ & $(193, 213)$ & $1.0$ & $0.975$ & BPR353 & $0.29$ & $0.29$ & $444$ & $4.9$ & $2.7$ & $0.51$ & $3.4$ & $0.76$ & $0.69$ & $0$ & $1.0$ & $26.6$\\
H& $0.8$ & $120$ & $(257, 256)$ & $1.0$ & $0.990$ & BPR353 & $0.20$ & $0.37$ & $514$ & $4.3$ & $14.1$ & $0.74$ & $1.8$ & $0.69$ & $0.80$ & $40$ & $4.5$ & $28.1$\\
& $0$ & $270$ & $(320, 256)^{\mathrm{f}}$ & $1.0$ & None & BPR353 & $0.47$ & $0.95$ & $685$ & $7.2$ & $5.5$ & $0.54$ & $7.9$ & $0.73$ & $0.33$ & $0$ & $1.0$ & $36.2$\\
& $0$ & $400$ & $(380, 256)^{\mathrm{f}}$ & $1.0$ & None & BPR353 & $0.16$ & $0.20$ & $880$ & $8.7$ & $5.4$ & $0.49$ & $14.5$ & $0.74$ & $0.24$ & $0$ & $1.0$ & $41.0$\\
& $46$ & $0$ & $(109, 170)$ & $1.0$ & $0.940$ & BPR353 & $0.38$ & $0.90$ & $763$ & $5.5$ & $23.5$ & $0.75$ & $2.4$ & $0.71$ & $0.77$ & $100$ & $7.3$ & $1.0$\\
H& $1.7$ & $240$ & $(321, 256)$ & $1.0$ & $0.992$ & BPR353 & $0.17$ & $0.28$ & $760$ & $6.7$ & $18.2$ & $0.70$ & $2.2$ & $0.70$ & $0.75$ & $44$ & $6.1$ & $35.3$\\
H& $3.4$ & $480$ & $(384, 341)^{\mathrm{f}}$ & $0.5$ & None & BPR353 & $0.46$ & $0.67$ & $580$ & $10.5$ & $9.2$ & $0.65$ & $1.5$ & $0.71$ & $0.77$ & $46$ & $8.0$ & $44.8$  &  (*)\\
& $100$ & $0$ & $(129, 213)$ & $1.0$ & $0.932$ & BPR353 & $0.26$ & $1.29$ & $1213$ & $10$ & $21.7$ & $0.67$ & $0.8$ & $0.72$ & $0.71$ & $100$ & $10.5$ & $1.0$\\
H& $6.8$ & $960$ & $(380, 256)^{\mathrm{f}}$ & $0.5$ & None & BPR353 & $1.54$ & $1.88$ & $843$ & $15.3$ & $10.3$ & $0.60$ & $1.9$ & $0.73$ & $0.69$ & $47$ & $11.0$ & $55.3$ & $\mathrm{^{(x)}}$\\
& $300$ & $0$ & $(129, 256)$ & $0.5$ & $0.962$ & BPR353 & $0.30$ & $1.22$ & $1235$ & $17.4$ & $12.4$ & $0.58$ & $19.2$ & $0.65$ & $0.23$ & $100$ & $16.7$ & $1.0$\\
H& $13$ & $1900$ & $(408, 256)^{\mathrm{f}}$ & $0.5$ & None & PC2 & $0.16$ & $0.24$ & $1252$ & $19.7$ & $13.8$ & $0.57$ & $21.2$ & $0.68$ & $0.20$ & $47$ & $14.2$ & $68.2$ & (**)\\
& $600$ & $0$ & $(129, 256)$ & $0.5$ & $0.962$ & BPR353 & $0.22$ & $0.43$ & $1640$ & $21.6$ & $21.4$ & $0.61$ & $11.6$ & $0.65$ & $0.32$ & $100$ & $20.9$ & $1.0$\\
\hline
\multicolumn{20}{c}{$E = 3 \times 10^{-5},\quad Pr = 0.3,\quad Sc = 3$} \\\hline
& $0$ & $0.5$ & $(49, 106)$ & $2.0$ & $0.779$ & CNAB2 & $1.37$ & $1.55$ & $113$ & $1.2$ & $0.4$ & $0.30$ & $52.3$ & $0.89$ & $0.86$ & $0$ & $1.0$ & $4.0$\\
H& $0.015$ & $0.5$ & $(49, 133)$ & $2.0$ & $0.779$ & CNAB2 & $1.95$ & $2.33$ & $154$ & $1.3$ & $1.1$ & $0.47$ & $24.0$ & $0.93$ & $0.90$ & $44$ & $1.3$ & $4.6$\\
H& $0.1$ & $1.1$ & $(49, 133)$ & $1.0$ & $0.779$ & CNAB2 & $2.15$ & $4.39$ & $98$ & $2$ & $0.4$ & $0.31$ & $36.5$ & $0.88$ & $0.88$ & $21$ & $1.4$ & $6.3$\\
H& $0.1$ & $1.1$ & $(129, 213)$ & $2.0$ & $0.960$ & BPR353 & $0.15$ & $3.11$ & $178$ & $1.7$ & $1.6$ & $0.48$ & $10.7$ & $0.85$ & $0.86$ & $22$ & $1.4$ & $6.4$\\
& $0.1$ & $1.4$ & $(129, 170)$ & $2.0$ & $0.960$ & BPR353 & $0.35$ & $1.79$ & $193$ & $1.8$ & $2.0$ & $0.51$ & $13.4$ & $0.83$ & $0.89$ & $19$ & $1.5$ & $7.2$\\
& $0$ & $2$ & $(65, 133)$ & $2.0$ & $0.848$ & CNAB2 & $1.81$ & $2.26$ & $204$ & $2.2$ & $1.6$ & $0.43$ & $8.5$ & $0.79$ & $0.86$ & $0$ & $1.0$ & $8.2$\\
& $0.5$ & $0$ & $(49, 133)$ & $2.0$ & $0.779$ & CNAB2 & $1.62$ & $1.71$ & $255$ & $1$ & $6.6$ & $0.68$ & $8.9$ & $0.96$ & $0.88$ & $100$ & $1.7$ & $1.0$\\
& $0$ & $4$ & $(97, 133)$ & $2.0$ & $0.975$ & CNAB2 & $1.24$ & $1.42$ & $288$ & $3.3$ & $2.0$ & $0.38$ & $2.3$ & $0.78$ & $0.60$ & $0$ & $1.0$ & $11.2$\\
H & $0.2$ & $4$ & $(97, 133)$ & $2.0$ & $0.930$ & CNAB2 & $1.04$ & $1.32$ & $307$ & $2.9$ & $4.4$ & $0.54$ & $3.4$ & $0.78$ & $0.81$ & $20$ & $2.0$ & $11.5$\\
& $1$ & $0$ & $(49, 106)$ & $1.0$ & $0.779$ & CNAB2 & $5.96$ & $6.41$ & $180$ & $1.9$ & $4.1$ & $0.69$ & $10.8$ & $0.92$ & $0.93$ & $100$ & $2.1$ & $1.0$\\
& $1$ & $0$ & $(49, 106)$ & $2.0$ & $0.864$ & CNAB2 & $0.79$ & $1.18$ & $281$ & $1.4$ & $20.6$ & $0.81$ & $3.4$ & $0.81$ & $0.78$ & $100$ & $2.4$ & $1.0$\\
& $0$ & $8$ & $(97, 133)$ & $2.0$ & $0.975$ & CNAB2 & $1.91$ & $5.45$ & $385$ & $4.6$ & $3$ & $0.41$ & $2.6$ & $0.77$ & $0.67$ & $0$ & $1.0$ & $14.8$\\
& $0.1$ & $1.1$ & $(97, 133)$ & $1.0$ & $0.935$ & CNAB2 & $1.30$ & $1.77$ & $164$ & $2.0$ & $9.2$ & $0.79$ & $5.0$ & $0.77$ & $0.80$ & $86$ & $2.2$ & $9.2$\\
H& $0.5$ & $5.5$ & $(129, 170)$ & $2.0$ & $0.960$ & CNAB2 & $1.21$ & $1.21$ & $373$ & $3.3$ & $8.4$ & $0.62$ & $2.3$ & $0.76$ & $0.81$ & $35$ & $2.4$ & $13.4$\\
& $0.1$ & $1.1$ & $(97, 133)$ & $2.0$ & $0.935$ & CNAB2 & $1.61$ & $1.89$ & $317$ & $1.8$ & $27.8$ & $0.79$ & $5.9$ & $0.75$ & $0.65$ & $87$ & $2.3$ & $8.9$\\
& $0.05$ & $5.5$ & $(65, 133)$ & $2.0$ & $0.864$ & CNAB2 & $1.63$ & $1.81$ & $342$ & $3$ & $12.8$ & $0.69$ & $2.2$ & $0.72$ & $0.80$ & $39$ & $2.2$ & $12.9$\\
& $1$ & $11$ & $(97, 133)$ & $1.0$ & $0.935$ & CNAB2 & $1.10$ & $1.30$ & $297$ & $5.7$ & $3.5$ & $0.51$ & $4.0$ & $0.78$ & $0.82$ & $39$ & $3.1$ & $17.9$\\
& $1$ & $11$ & $(145, 170)$ & $2.0$ & $0.960$ & CNAB2 & $1.08$ & $1.23$ & $541$ & $4.9$ & $11.4$ & $0.59$ & $1.5$ & $0.75$ & $0.76$ & $39$ & $3.2$ & $17.4$\\
& $0$ & $20$ & $(97, 133)$ & $2.0$ & $0.975$ & CNAB2 & $1.04$ & $1.06$ & $581$ & $6.7$ & $5.5$ & $0.43$ & $5.6$ & $0.76$ & $0.44$ & $0$ & $1.0$ & $20.0$\\
& $3.7$ & $0$ & $(49, 133)$ & $1.0$ & $0.779$ & CNAB2 & $1.31$ & $1.41$ & $328$ & $5$ & $12.4$ & $0.71$ & $2.8$ & $0.74$ & $0.84$ & $100$ & $4.0$ & $1.0$\\
& $3.7$ & $0$ & $(81, 133)$ & $2.0$ & $0.900$ & CNAB2 & $0.88$ & $1.02$ & $613$ & $4.4$ & $36.7$ & $0.72$ & $3.7$ & $0.73$ & $0.70$ & $100$ & $4.0$ & $1.0$\\
& $1.7$ & $20$ & $(145, 170)$ & $2.0$ & $0.970$ & CNAB2 & $0.83$ & $1.08$ & $738$ & $7.1$ & $12.5$ & $0.54$ & $1.7$ & $0.76$ & $0.65$ & $40$ & $4.0$ & $21.3$\\
& $5$ & $0$ & $(97, 133)$ & $2.0$ & $0.864$ & CNAB2 & $0.89$ & $1.06$ & $726$ & $5.7$ & $38.6$ & $0.69$ & $2.8$ & $0.72$ & $0.71$ & $100$ & $4.6$ & $1.0$\\
& $2.3$ & $27$ & $(129, 133)$ & $2.0$ & $0.960$ & CNAB2 & $1.07$ & $1.07$ & $857$ & $8.3$ & $14.7$ & $0.53$ & $1.7$ & $0.76$ & $0.63$ & $41$ & $4.6$ & $23.3$\\
& $0.2$ & $30$ & $(109, 256)$ & $2.0$ & $0.948$ & BPR353 & $1.02$ & $1.10$ & $837$ & $8.4$ & $20$ & $0.57$ & $1.5$ & $0.73$ & $0.70$ & $40$ & $4.0$ & $23.0$\\
H& $0.4$ & $8$ & $(97, 170)$ & $2.0$ & $0.962$ & BPR353 & $0.95$ & $1.09$ & $755$ & $6.9$ & $42.4$ & $0.68$ & $4.2$ & $0.69$ & $0.63$ & $84$ & $4.2$ & $17.3$\\
& $3$ & $35$ & $(129, 170)$ & $2.0$ & $0.962$ & CNAB2 & $1.13$ & $1.13$ & $976$ & $9.5$ & $16.4$ & $0.52$ & $1.8$ & $0.76$ & $0.59$ & $42$ & $5.2$ & $25.2$  &  \\
H& $0.3$ & $35$ & $(129, 213)$ & $2.0$ & $0.962$ & BPR353 & $0.43$ & $0.51$ & $949$ & $9.7$ & $24.5$ & $0.57$ & $1.6$ & $0.73$ & $0.67$ & $48$ & $4.6$ & $24.2$  &  \\
& $0$ & $100$ & $(193, 213)$ & $2.0$ & $0.970$ & BPR353 & $0.13$ & $0.18$ & $1378$ & $12.5$ & $12.4$ & $0.40$ & $26.0$ & $0.68$ & $0.16$ & $0$ & $1.0$ & $31.5$\\
& $12$ & $0$ & $(129, 170)$ & $2.0$ & $0.960$ & CNAB2 & $0.94$ & $1.02$ & $1237$ & $11$ & $37.5$ & $0.59$ & $2.5$ & $0.73$ & $0.66$ & $100$ & $6.8$ & $1.0$\\
& $7$ & $70$ & $(193, 170)$ & $2.0$ & $0.983$ & CNAB2 & $0.87$ & $1.22$ & $1579$ & $13.9$ & $19$ & $0.44$ & $27.2$ & $0.69$ & $0.15$ & $47$ & $7.2$ & $31.1$\\
& $9$ & $100$ & $(193, 256)$ & $2.0$ & $0.983$ & PC2 & $0.10$ & $0.30$ & $1765$ & $14.8$ & $27$ & $0.49$ & $12.2$ & $0.72$ & $0.27$ & $45$ & $8.2$ & $33.1$\\
H& $0.9$ & $100$ & $(193, 213)$ & $2.0$ & $0.983$ & PC2 & $0.12$ & $0.16$ & $1790$ & $17.4$ & $26.5$ & $0.47$ & $26.3$ & $0.71$ & $0.15$ & $51$ & $7.3$ & $32.6$\\
& $40$ & $0$ & $(97, 213)$ & $2.0$ & $0.960$ & CNAB2 & $0.34$ & $0.38$ & $2521$ & $19.5$ & $50$ & $0.50$ & $22.2$ & $0.71$ & $0.17$ & $100$ & $11.0$ & $1.0$\\
& $68$ & $0$ & $(129, 170)$ & $1.0$ & $0.960$ & BPR353 & $0.59$ & $0.59$ & $1671$ & $23.5$ & $33$ & $0.53$ & $7.4$ & $0.71$ & $0.26$ & $100$ & $13.0$ & $1.0$\\
\hline
\multicolumn{20}{c}{$E = 1 \times 10^{-4},\quad Pr = 0.3,\quad Sc = 3$} \\\hline
& $0.013$ & $0.045$ & $(49, 85)$ & $5.0$ & None & CNAB2 & $1.09$ & $1.78$ & $175$ & $1.1$ & $1.6$ & $0.32$ & $10.1$ & $0.98$ & $0.82$ & $40$ & $1.2$ & $2.8$\\
& $0$ & $0.1$ & $(65, 106)$ & $5.0$ & None & CNAB2 & $1.17$ & $1.31$ & $161$ & $1.3$ & $3.2$ & $0.47$ & $4.9$ & $0.81$ & $0.88$ & $0$ & $1.0$ & $3.8$\\
& $0.02$ & $0.09$ & $(49, 85)$ & $5.0$ & None & CNAB2 & $2.54$ & $7.73$ & $180$ & $1.2$ & $7.2$ & $0.62$ & $4.3$ & $0.86$ & $0.86$ & $36$ & $1.3$ & $4.0$\\
& $0.1$ & $0$ & $(65, 106)$ & $5.0$ & None & CNAB2 & $1.86$ & $2.16$ & $363$ & $1.3$ & $34.7$ & $0.67$ & $11.9$ & $0.90$ & $0.63$ & $100$ & $1.7$ & $1.0$\\
& $0$ & $0.5$ & $(49, 85)$ & $5.0$ & None & CNAB2 & $1.65$ & $1.99$ & $378$ & $3.9$ & $4$ & $0.32$ & $1.8$ & $0.79$ & $0.58$ & $0$ & $1.0$ & $7.6$\\
& $0.047$ & $0.33$ & $(65, 85)$ & $5.0$ & None & BPR353 & $0.50$ & $3.64$ & $345$ & $2.6$ & $14.8$ & $0.58$ & $2.8$ & $0.78$ & $0.75$ & $37$ & $1.6$ & $6.9$\\
& $0.08$ & $0.19$ & $(65, 106)$ & $5.0$ & None & CNAB2 & $1.10$ & $1.18$ & $334$ & $1.8$ & $31$ & $0.69$ & $3.8$ & $0.81$ & $0.71$ & $67$ & $1.7$ & $6.0$\\
& $0$ & $0.7$ & $(49, 85)$ & $5.0$ & None & CNAB2 & $2.02$ & $2.43$ & $438$ & $4.7$ & $4.8$ & $0.31$ & $2.8$ & $0.78$ & $0.52$ & $0$ & $1.0$ & $8.6$\\
& $0.07$ & $0.5$ & $(97, 106)$ & $5.0$ & None & BPR353 & $0.35$ & $1.70$ & $438$ & $3.4$ & $18.9$ & $0.56$ & $3.0$ & $0.78$ & $0.71$ & $40$ & $1.8$ & $8.1$\\
& $0$ & $1$ & $(65, 106)$ & $5.0$ & None & CNAB2 & $1.06$ & $1.33$ & $522$ & $5.6$ & $5.5$ & $0.31$ & $8.1$ & $0.77$ & $0.35$ & $0$ & $1.0$ & $9.9$\\
& $0.14$ & $0.34$ & $(65, 106)$ & $5.0$ & None & CNAB2 & $1.08$ & $1.09$ & $482$ & $3.1$ & $38.6$ & $0.64$ & $3.8$ & $0.78$ & $0.68$ & $70$ & $2.0$ & $7.7$\\
& $0.19$ & $0$ & $(65, 106)$ & $5.0$ & None & CNAB2 & $1.14$ & $1.36$ & $469$ & $2.3$ & $60.1$ & $0.69$ & $5.0$ & $0.80$ & $0.64$ & $100$ & $2.0$ & $1.0$\\
& $0.12$ & $0.52$ & $(81, 106)$ & $5.0$ & None & BPR353 & $0.89$ & $2.24$ & $516$ & $3.8$ & $27.1$ & $0.58$ & $3.3$ & $0.78$ & $0.69$ & $56$ & $2.0$ & $8.6$\\
& $0.027$ & $1.1$ & $(65, 106)$ & $5.0$ & None & CNAB2 & $2.11$ & $2.44$ & $580$ & $5.9$ & $7.7$ & $0.34$ & $6.3$ & $0.78$ & $0.36$ & $10$ & $1.9$ & $10.4$\\
& $0.027$ & $1.1$ & $(65, 106)$ & $5.0$ & None & CNAB2 & $1.14$ & $1.30$ & $570$ & $5.7$ & $8.8$ & $0.37$ & $2.9$ & $0.79$ & $0.54$ & $11$ & $1.9$ & $10.3$\\
& $0.1$ & $0.69$ & $(65, 106)$ & $5.0$ & None & CNAB2 & $1.12$ & $1.45$ & $533$ & $4.2$ & $22.8$ & $0.55$ & $3.4$ & $0.78$ & $0.69$ & $44$ & $2.0$ & $9.1$\\
& $0.049$ & $1$ & $(65, 106)$ & $5.0$ & None & CNAB2 & $1.05$ & $1.16$ & $573$ & $5.4$ & $11.1$ & $0.41$ & $2.5$ & $0.79$ & $0.60$ & $20$ & $1.9$ & $10.1$\\
& $0.049$ & $1$ & $(65, 133)$ & $5.0$ & None & BPR353 & $0.13$ & $1.38$ & $566$ & $5.4$ & $11.2$ & $0.42$ & $2.2$ & $0.78$ & $0.61$ & $20$ & $1.9$ & $10.1$\\
& $0$ & $1.4$ & $(65, 106)$ & $5.0$ & None & CNAB2 & $1.32$ & $1.61$ & $615$ & $6.5$ & $7$ & $0.31$ & $16.6$ & $0.75$ & $0.23$ & $0$ & $1.0$ & $11.0$\\
& $0.14$ & $1$ & $(65, 106)$ & $3.0$ & None & BPR353 & $0.33$ & $2.10$ & $412$ & $5.7$ & $11.8$ & $0.50$ & $2.5$ & $0.79$ & $0.74$ & $45$ & $2.3$ & $10.6$\\
& $0.14$ & $1$ & $(81, 106)$ & $5.0$ & None & BPR353 & $0.37$ & $1.72$ & $655$ & $5.5$ & $24.7$ & $0.51$ & $3.3$ & $0.77$ & $0.66$ & $46$ & $2.3$ & $10.5$\\
& $0.14$ & $1$ & $(81, 133)$ & $7.0$ & $0.909$ & BPR353 & $0.42$ & $2.36$ & $896$ & $5.2$ & $40.6$ & $0.52$ & $4.3$ & $0.77$ & $0.62$ & $46$ & $2.3$ & $10.4$\\
& $0$ & $2$ & $(97, 170)$ & $5.0$ & $0.935$ & BPR353 & $0.88$ & $2.98$ & $750$ & $7.2$ & $8.7$ & $0.32$ & $39.4$ & $0.72$ & $0.10$ & $0$ & $1.0$ & $12.4$\\
& $0.3$ & $0$ & $(61, 170)$ & $5.0$ & None & CNAB2 & $0.16$ & $1.24$ & $616$ & $3.6$ & $72.6$ & $0.66$ & $4.7$ & $0.76$ & $0.62$ & $100$ & $2.4$ & $1.0$\\
& $0.21$ & $1.5$ & $(81, 133)$ & $5.0$ & $0.909$ & BPR353 & $0.32$ & $2.74$ & $813$ & $6.8$ & $32$ & $0.51$ & $3.9$ & $0.77$ & $0.63$ & $49$ & $2.7$ & $12.0$\\
& $0.3$ & $3$ & $(97, 170)$ & $5.0$ & $0.935$ & BPR353 & $0.28$ & $2.13$ & $1144$ & $9.6$ & $30.2$ & $0.44$ & $6.8$ & $0.79$ & $0.44$ & $43$ & $3.3$ & $14.6$\\
& $0.9$ & $0$ & $(81, 170)$ & $5.0$ & $0.909$ & BPR353 & $0.34$ & $2.08$ & $1244$ & $9.8$ & $74.9$ & $0.54$ & $6.9$ & $0.74$ & $0.56$ & $100$ & $3.8$ & $1.0$\\
& $0.4$ & $6$ & $(97, 213)$ & $5.0$ & $0.935$ & BPR353 & $0.29$ & $1.85$ & $1576$ & $12.8$ & $30.6$ & $0.38$ & $42.7$ & $0.72$ & $0.09$ & $35$ & $4.0$ & $17.4$\\
& $5$ & $0$ & $(73, 213)$ & $5.0$ & $0.889$ & BPR353 & $0.33$ & $1.49$ & $3363$ & $23.5$ & $130$ & $0.39$ & $17.7$ & $0.75$ & $0.25$ & $100$ & $7.4$ & $1.0$\\
& $20$ & $0$ & $(129, 170)$ & $5.0$ & $0.960$ & CNAB2 & $0.12$ & $0.39$ & $6160$ & $45.1$ & $297$ & $0.36$ & $21.9$ & $0.76$ & $0.21$ & $100$ & $10.5$ & $1.0$
\end{longtable}
}

%% file: appendix_simu.tex
\section{Simulation $^\mathrm{x}$}
\label{exception}
\begin{figure}
	\centering
	\includegraphics[width=0.45\textwidth]{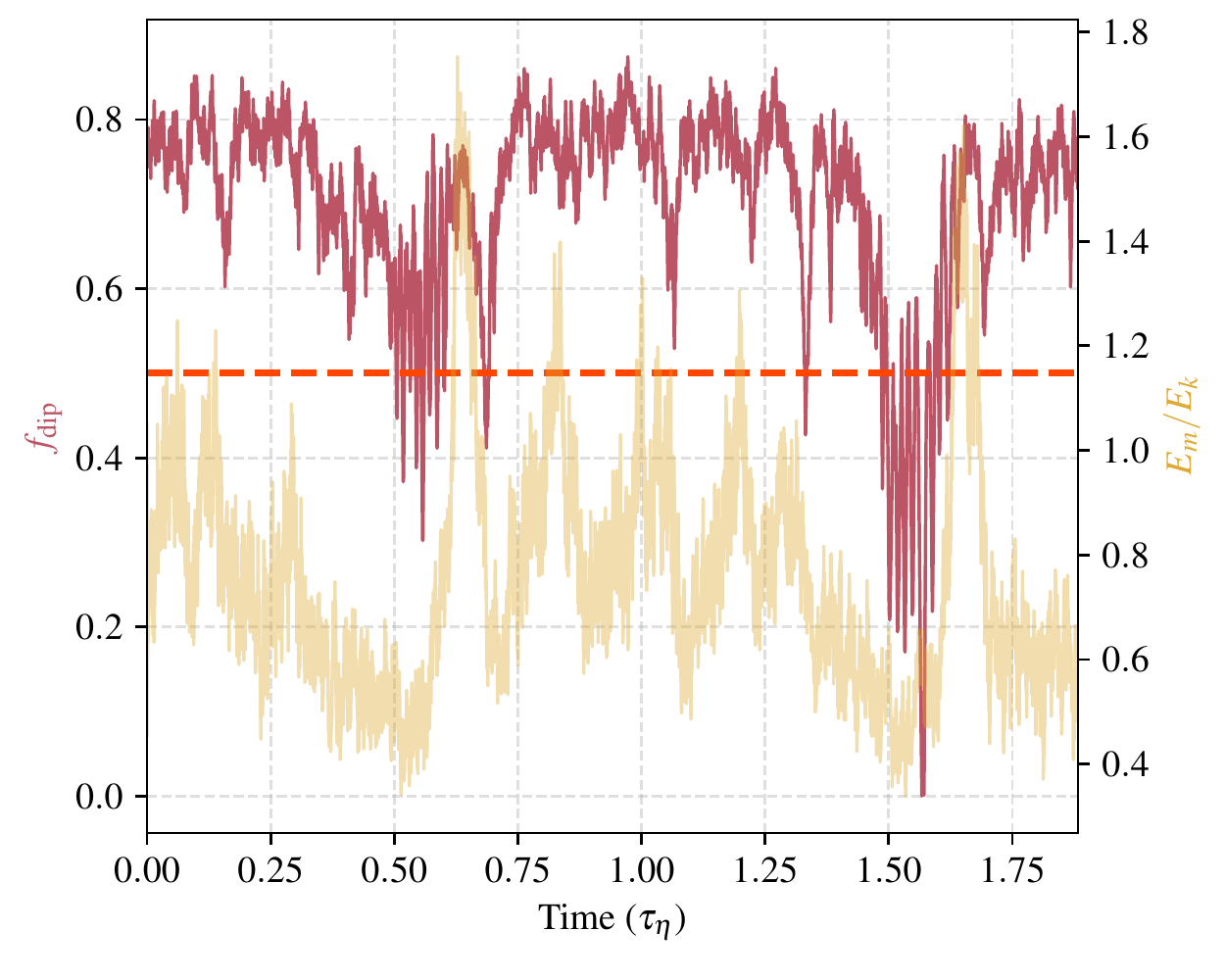}
	\caption{Time evolution of the dipolar fraction $\fdip$ 
		and of the magnetic to kinetic energy ratio $E_m/E_k$ 
		for the simulation ($\mathrm{x}$) of 
		Table~\ref{simu_tab}. The horizontal dashed 
		line corresponds to the boundary between dipole-dominated and 
		multipolar dynamos ($\fdip=0.5$). Time is scaled by the magnetic 
	        diffusion time.}
	\label{fdip_68}
\end{figure}

In this appendix we provide in Figure~\ref{fdip_68} the detailed time 
evolution of the dipolar fraction $\fdip$ and of
the magnetic to kinetic energy ratio $E_m/E_k$ of the anomalous simulation that appears for instance in the top-right
quadrant of Fig.~\ref{menu}(b) and Fig.~\ref{ek_em}. This simulation is marked 
with a ${(\mathrm{x})}$ in Table~\ref{simu_tab}. 

